\documentclass[10pt]{article} 
\usepackage{amsmath, amsfonts, amsthm, amssymb, amscd}
\usepackage[mathcal]{euscript} 
\usepackage{dsfont, pxfonts, mathrsfs, verbatim}  
\theoremstyle{plain}
\newtheorem{theorem}{Theorem}[section]
\newtheorem{proposition}[theorem]{Proposition} 
\newtheorem{lemma}[theorem]{Lemma}
 
\newtheorem{definition}[theorem]{Definition} 
\newtheorem{remark}[theorem]{Remark}  
\numberwithin{equation}{section} 
\usepackage{a4wide}
\newcommand{\auth}{\textsc} 
\newcommand{\jou}{\textit}

\newcommand \Pit {\widetilde \Pi} 
\newcommand \Rp {R'}

\newcommand \Rmax  {R_\text{max}}
\newcommand \conf  {\text{conf}}

\newcommand \Ubz  {{\mathop{\overline U}_{0}}}
\newcommand \Abz  {{\mathop{\overline A}_{0}}}
\newcommand \nubz  {{\mathop{\overline \nu}_{0}}}
\newcommand \etabz  {{\mathop{\overline \eta}_{0}}}
\newcommand \Rbz  {{\mathop{\overline R}_{0}}}

\newcommand \Sigmat {\Sigma} 
\newcommand \Sigmah {\Sigma}  
\newcommand \Omegat {\widetilde \Omega} 
 
\newcommand \muh M 
\newcommand \vv V

\newcommand \Fcal {\mathcal F} 
\newcommand \Gcal {\mathcal G}  
\newcommand \Ecal {\mathcal E}  
\newcommand \rhob {\overline \rho}
\newcommand \nub {\overline \nu}
\newcommand \etab {\overline \eta}

\newcommand \Ub {\overline U}
\newcommand \Ab {\overline A}
\newcommand \Mb {\overline M}
\newcommand \Vb {\overline V}
\newcommand \gb {\overline g}
\newcommand \kb {\overline k}
\newcommand \jb {\overline j}
\newcommand \Rb {\overline R}
\newcommand \alphab {\overline \alpha \, }
\newcommand \nablab {\overline \nabla} 
\newcommand \jtn {\widehat j}  
\newcommand \Stn {\check {S}}  
\newcommand \rhot {\widetilde \rho} 
\newcommand \mut {\widetilde \mu} 
\newcommand \pt {\widetilde p} 
\newcommand \jt {\widetilde j}  
\newcommand \St {\widetilde S} 
\newcommand \gthree {{}^{(3)}g} 
\newcommand \gtwo {{}^{(2)}g} 
\newcommand \tr {\text{tr }}
\newcommand \loc   {\text{loc}}

\newcommand \RR 		{\mathbb{R}}  
\newcommand \del 		\partial
\newcommand \eps 		\epsilon
\newcommand \lam 		\lambda 
\newcommand \be 		{\begin{equation}}
\newcommand \ee 		{\end{equation}}

\newcommand \Lcal 	 	{\mathcal L}
\newcommand \Dcal 	 	{\mathcal D}

\begin{document} 

\title 
{A global foliation of Einstein--Euler spacetimes
\\
with Gowdy-symmetry on $T^3$ 
}
%      Correct spelling:   LeFloch  or LeFLOCH
\author{Philippe G. LeFloch\footnote{
Laboratoire Jacques-Louis Lions \& Centre National de la Recherche Scientifique, 
Universit\'e Pierre et Marie Curie (Paris 6), 4 Place Jussieu, 75252 Paris, France. 
\newline 
E-mail: {\tt pgLeFloch@gmail.com.} 
Web: {\tt philippelefloch.wordpress.com.} 
} 
\, and 
Alan D. Rendall\footnote{Max-Planck-Institut f\"ur Gravitationsphysik, 
Albert-Einstein Institut, Am M\"uhlenberg 1, 14476 Potsdam, Germany. 
E-mail: {\tt Rendall@aei.mpg.de.}
\newline Published in: \tt Archive for Rational Mechanics and Analysis (2011).  
}
}  
\date{} 
\maketitle

\begin{abstract} We investigate the initial value problem for the Einstein--Euler equations
of general relativity under the assumption of Gowdy symmetry on $T^3$, and we construct 
matter spacetimes with low regularity. 
These spacetimes admit, both, impulsive gravitational waves in the metric  
(for instance, Dirac mass curvature singularities propagating at light speed)  
and shock waves in the fluid (i.e., discontinuities propagating at about the sound speed). 
Given an initial data set, we establish the existence of a future development 
and we provide a global foliation in terms of a globally and geometrically defined time-function, 
closely related to the area of the orbits of the symmetry group. 
The main difficulty lies in the low regularity assumed on the initial data set 
which requires a distributional formulation of the Einstein--Euler equations.  
\end{abstract} 
 
%-----------------------------------------------------------------------------------------------------------------

\section{Introduction}
\label{IN}
  
\subsection*{Background}
  
We consider matter spacetimes with $T^3$ Gowdy symmetry which, by definition, admit a two-parameter group of 
isometries generated by two orthogonally transitive, commuting Killing fields.
(See Section~\ref{Main}, below.) 
Under this symmetry assumption,  
the initial value problem for the Einstein equations has received much attention in recent years, 
both in the vacuum case and in the matter case when the matter is governed by the Vlasov equation of 
the kinetic theory of gases. In the present paper, we are interested in the evolution of {\sl perfect fluids,}  
and in the context of Gowdy symmetry, 
we aim at constructing a globally foliated, future development of a given initial data set.
The main novelty in this work lies in the presence of {\sl shock waves} which appear in the fluid and, in turn, 
generate curvature discontinuities propagating in the spacetime.  
 
Recall that a spacetime is a $(3+1)$-dimensional differential manifold $M$ that is
endowed with a Lorentzian metric $g$ with signature $(-,+,+,+)$, satisfying the Einstein field equations 
\be
\label{Einstein}
G_{\alpha\beta} = \kappa \, T_{\alpha\beta}, 
\ee
where $\kappa>0$ is a constant and all Greek indices lie in the range ${0, \ldots, 3}$. 
Here, $G_{\alpha\beta} := R_{\alpha\beta} - ({R_\gamma}^\gamma/2) g_{\alpha\beta}$ denotes Einstein's curvature tensor, 
$R_{\alpha\beta}$ the Ricci curvature, and ${R_\gamma}^\gamma$ the scalar curvature of the manifold. 
The stress-energy tensor $T_{\alpha\beta}$ appearing in the right-hand side of \eqref{Einstein}
describes the matter content of the spacetime which, for perfect fluids, reads 
\be
\label{momentum}
T_{\alpha\beta} := (\mu+p) \, u_\alpha u_\beta + p \, g_{\alpha\beta},
\ee 
where the scalar field $\mu$ represents the mass-energy density of the fluid and $u$ its velocity vector. 
The spacetime is assumed to be time-oriented and $u$ is normalized to be 
a future-oriented, unit timelike vector field. We also assume the linear density-pressure law 
\be
\label{pressure}
p := k^2 \, \mu,  
\ee
in which the constant $k \in (0,1)$ represents the sound speed in the fluid 
and does not exceed the light speed normalized to be $1$. 

Under the assumption of $T^3$ Gowdy symmetry made in the present paper
and after introducing areal or conformal coordinates (see Section~\ref{EE}, below), 
the Einstein field equations take the form of a coupled system of nonlinear wave equations with differential constraints. 
Since the pioneering work by Gowdy \cite{Gowdy}, (vacuum) Gowdy symmetric spacetimes have been extensively 
studied  \cite{Moncrief,EardleyMoncrief,IsenbergMoncrief,Chrusciel,ChruscielIsenbergMoncrief,BergerChruscielMoncrief,BergerChruscielIsenbergMoncrief,IsenbergWeaver} and 
Penrose's strong cosmic censorship conjecture \cite{Penrose,HawkingPenrose} 
was eventually established by Ringstr\"om \cite{Ringstrom1,Ringstrom2}. 
A generalization of these spacetimes that contain matter governed by the Vlasov equation was recently presented 
in \cite{Rendall1,Rendall2,Andreasson,AndreassonRendallWeaver,DafermosRendall,Smulevici}. 

As far as compressible matter is concerned, the mathematical investigation of Gowdy-type spacetimes
 was initiated by 
LeFloch and Stewart \cite{LeFlochStewart1} (see also \cite{BLSS}), 
who introduced a converging approximation scheme for the initial value problem and derived several 
a~priori bounds in suitably chosen coordinates.
Therein, it was found necessary to cope with weak solutions to the Einstein equations, 
understood in the distributional sense and containing propagating discontinuities (shock waves).
In \cite{BLSS,LeFlochStewart1}, the authors established a {\sl local-in-time} existence result 
in the class of spacetimes with bounded variation.  
The present work is a continuation of this work and is aimed at constructing a 
{\sl global} foliation of such spacetimes.

Recall that spacetimes with bounded variation were constructed by 
Christo\-doulou in his work \cite{Christodoulou1} that settled positively the weak version of Penrose's 
cosmic censorship conjecture in the context of spherically symmetric spacetimes and for scalar fields.
Recall also that Groah and Temple \cite{GroahTemple}
established a {\sl local-in-time} existence result for spherically symmetric matter spacetimes. 
In such spacetimes, no gravitational waves are permitted and the matter equations are coupled with a 
differential equation accounting for (non-evolutive) geometrical features.

The study of solutions of low regularity is motivated by the fact that these
can arise from smooth initial data. In the case of the Euler equations without gravity,
this is well known and it is physically plausible that adding
gravity should not make a fundamental difference. A proof that this loss of regularity occurs in general relativity was given in \cite{rendallstahl}. That result concerned plane symmetric solutions of the Einstein--Euler equations, and they are a special case of the solutions studied in what follows. A direct comparison is made difficult by the fact the time coordinate used in
\cite{rendallstahl}, a constant mean curvature coordinate, is different from that used in the present paper.

%-------------------------------------------------------------------------------------------------------------

\subsection*{Objectives of this paper}

The assumption of Gowdy symmetry made in the present paper allows us to address 
the properties of dynamical gravitational degrees of freedom. 
We consider the initial value problem, and search for the spacetime as a future development of a 
prescribed initial data set. 
This amounts to prescribing an initial hypersurface endowed with a Riemannian metric 
and its second fundamental form, together with the mass-energy density and current vector determined by  
the fluid on this hypersurface.
Our main result provides {\sl global} existence for the Einstein--Euler equations
within a class of  
spacetimes with low regularity which may contain, both, impulsive gravitational waves and shock waves:
\begin{itemize}
\item When a {\sl shock wave} arises in the fluid, the fluid variables become discontinuous and, as a consequence of 
Einstein's equations, the spacetime curvature becomes discontinuous.
\item In addition, our theory allows
for distributional curvature singularities propagating at light speed. Such waves are referred to as 
{\sl impulsive gravitational waves.}  
\item  
Our framework also encompasses the situation where the fluid contains low density regions, even vacuum regions. 
\end{itemize}
The theorem established in this paper 
can be regarded as a statement on the {\sl nonlinear stability} of (vacuum) spacetimes with Gowdy symmetry on $T^3$ when compressible matter is included. Indeed, our matter spacetimes can be made to be arbitrarily close (in a well-defined functional norm) to vacuum spacetimes, 
by choosing the total amount of matter energy on the initial hypersurface to be arbitrarily small.

More precisely, by combining geometrical and analytical arguments,  
we establish the existence of a {\sl global foliation}  
based on a geometrically defined time-function coinciding with the area of the orbits of the symmetry group.  
Our approach is motivated by pioneering works by Moncrief \cite{Moncrief} (Gowdy spacetimes) 
and Berger, Chru\'sciel, Isenberg, and Moncrief \cite{BergerChruscielIsenbergMoncrief} 
(vacuum spacetimes with $T^2$ symmetry) 
which established the existence of such a foliation for {\sl vacuum} spacetimes. 
However, we bring in a conceptually new and mathematically challenging aspect in that 
we consider solutions that have very low regularity.
In the class under consideration, many (high-order) estimates derived in \cite{BergerChruscielIsenbergMoncrief}
no longer hold and must be bypassed. 
The regularity of our spacetimes is considerably lower than 
the one constructed in earlier works and, for this reason, our analysis encompasses a {\sl larger} class   
of spacetimes.

Concerning the regularity of the spacetimes constructed here, the following features should be stressed:  
\begin{itemize}

\item {\sl Natural estimates for the geometry.} As we will show, it is natural to impose 
that the essential metric coefficients belong to the Sobolev space $H^1$ of functions which, by definition, 
are square-integrable, together with 
their first-order derivatives.  This regularity is dictated by the energy-type functional associated with the problem under consideration. 

\item {\sl Natural estimates for the fluid.}  
As we will also show, it is natural to impose that the energy density and the current vector 
belong to the Lebesgue space $L^1$ of integrable functions, 
but no further regularity can be imposed on their derivatives.  
Our setting allows the fluid variables to contain vacuum states 
and the essential velocity component to approach the speed of light. 

\end{itemize}

\noindent For further results on the existence and qualitative  
properties of spacetimes with matter when very low regularity is assumed, we refer to 
LeFloch and Stewart \cite{LeFlochStewart1, LeFlochStewart2}. 

An outline of this paper follows. In Section~\ref{Main}, we present our assumptions and main results; see Theorem~\ref{maintheo}. In Section~\ref{EE}, we express the Einstein--Euler equations first in a general foliation 
and, then, 
in areal and in conformal coordinates. Next, in both the expanding and the contracting directions 
handled in Sections~\ref{Expa} and \ref{contra}, respectively, 
we introduce suitable notions of weak solutions to the Einstein--Euler equations
(Definitions~\ref{finit-areal} and \ref{finit}) 
adapted to the setting under consideration, and we derive analytic and geometric properties. 
Finally, having successfully determined the natural weak regularity assumptions 
required on the initial data and enjoyed by the solutions to the Einstein--Euler system,  
we conclude with the compactness framework established in LeFloch~\cite{LeFloch} and presented in Section~\ref{Compac}
below. 

%==================================================================================================================

\section{Global foliations of matter spacetimes}
\label{Main}

\subsection*{Symmetry and regularity assumptions}

First, we need to specify the symmetry and regularity conditions characterizing the class of spacetimes $(M,g)$ 
under consideration. 
Throughout we consider matter spacetimes satisfying the Einstein--Euler equations \eqref{Einstein}--\eqref{pressure}. 
We assume that they admit an Abelian $T^2$ isometry group with spacelike orbits  
generated by two linearly independent, commuting, spacelike Killing fields $K,L$ 
whose twist constants vanish, that is, 
\be
\label{vani} 
c^K := \eps_{\alpha\beta\gamma\delta} K^\alpha L^\beta \nabla^\gamma K^\delta = 0, 
\qquad 
c^L := \eps_{\alpha\beta\gamma\delta} K^\alpha L^\beta \nabla^\gamma L^\delta = 0,  
\ee
where $\eps_{\alpha\beta\gamma\delta}$ denotes the totally anti-symmetric elementary tensor. 
This latter condition expresses the assumption that the frame of covectors $\big( g(K, \cdot), g(L, \cdot)\big)$
is integrable in the sense of Frobenius, and 
was identified by Chru\'sciel \cite{Chrusciel} in order to single out (vacuum) Gowdy spacetimes \cite{Gowdy} 
within the larger class of $T^2$ symmetric spacetimes. Note in passing 
that the scalars $c^K, c^L$ must be constant throughout any $T^2$ symmetric {\sl vacuum} spacetime, 
as was observed by Geroch \cite{Geroch1,Geroch2}.  
Matter spacetimes satisfying the conditions above are said to have {\sl Gowdy symmetry.}  
The topology of the manifold, denoted by $M$, must also be specified, and we assume here that the spatial 
slices have the $T^3$ topology. 
All of the above conditions determine the class of {\sl spacetimes with Gowdy symmetry on $T^3$,} 
under consideration in the present paper.
In the vacuum case, these are precisely the so-called Gowdy spacetimes first studied in \cite{Gowdy}. 
  
We are interested in the future Cauchy development of a given initial data set,  
and the global topology will turn out to be 
$M = [c_0, c_1) \times T^3$ for some $c_0 < c_1 \leq \infty$,
 and
 the spacetime to be foliated by spacelike hypersurfaces 
along which a time-function (denoted below by $t$ or $\tau$) remains constant.   
The main unknowns of the theory are the Lorentzian metric $g$ describing the geometry,  
together with the scalar field $\mu$ and the vector field $u$ characterizing the matter content
(via \eqref{momentum}). 
In local coordinates $x=(t, x^a)$ $(a=1,2,3$), Einstein's field equations \eqref{Einstein} will be decomposed
into evolution 
and constraint equations for the metric coefficients which
 should also be coupled with the Euler equations for the evolution of the fluid variables 
(see \eqref{Euler}, below). 
It will be convenient to express the Euler equations as evolution equations 
for the mass-energy density $\rho$ and the momentum $j$, measured by {\sl an observer moving orthogonally} 
to the hypersurfaces determined by the time-function. 
By construction, $j$ is tangent to the leaves of the foliation, and
prescribing the fields $\rho, j$ is equivalent to prescribing the fields $\mu, u$ (see \eqref{rhojS}, below).

We now discuss the regularity of the spacetimes under consideration. 
Our regularity assumptions must allow for propagating discontinuities
and, in view of the expression of the energy functional associated with the Einstein--Euler equations
(see Lemma~\ref{energybound}, below),  
it is natural to require that on each slice labeled by the parameter $t$, 
both fields $\rho(t, \cdot), j(t, \cdot)$ belong to the space $L^1(T^3)$.
In short, we write  
\be
\label{reg-fluid} 
\rho, \, j \in L^\infty_\loc ([c_0, c_1), L^1(T^3)).  
\ee 
All the Lebesgue and Sobolev spaces under consideration are 
endowed with the volume form induced  
by the Lorentzian metric. By definition, for almost every $t$ the function 
$\rho(t)$ is well-defined and integrable, and
the function $t \mapsto \|\rho(t, \cdot) \|_{L^1(T^3)}$ (i.e.~the integral of the function $\rho(t)$)
is bounded uniformly for almost all $t$ in any compact subinterval of $[c_0, c_1)$. 
As we show later, the regularity \eqref{reg-fluid} is sufficient to formulate the Euler equations in the sense of 
distributions. In addition, following the general theory of nonlinear hyperbolic systems \cite{Dafermos-book,LeFloch-book}, 
we impose that $\rho, j$ satisfy certain entropy inequalities (see \eqref{entr}, below)
which select the physically meaningful solutions to the Euler equations.

On the other hand, again in view of the expression of the energy functional  
we impose that on each slice the essential metric coefficients belong to the Sobolev space $H^1$.
In short, we write  
\be
\label{reg-metric} 
g \in L^\infty_\loc ([c_0, c_1), H^1(T^3)). 
\ee
It should be noted that our precise definition, given in Sections~\ref{Expa} and \ref{contra} below, 
are (slightly) more general and allow certain components of the metric to be {\sl less regular} than $H^1$.  
 
Due to the assumed symmetry, it will turn out that 
\eqref{reg-metric} implies that the metric coefficients remain locally bounded, that is, 
$g \in L^\infty_\loc$, and 
our regularity condition suffices to express Einstein's equations in a weak sense.
Indeed, the Riemann curvature of $g$ and all of its traces and, therefore, the Einstein tensor  
are then well-defined as distributions~\cite{LeFlochMardare}.  Note finally that the regularity 
described here may in principle depend on the foliation under consideration, 
and may not be as geometric as one may wish.

We refer to such a set $(M,g, \rho, j)$ having the regularity \eqref{reg-fluid}-\eqref{reg-metric} 
as a {\sl finite energy spacetime with Gowdy symmetry on $T^3$.} 

%-------------------------------------------------------------------------------------------------

\subsection*{The initial value problem}

To formulate the initial value problem we prescribe a Riemannian metric $\gb$ on $T^3$ together with a $2$-covariant
symmetric tensor field $\kb$, both of them satisfying the assumption of Gowdy symmetry on $T^3$. 
By definition, the Lie group $T^2$ acts as an isometry group on the torus $T^3$ generated by two 
(linearly independent, commuting) vector fields,  
$$ 
\Lcal_X \gb = \Lcal_Y \gb = 0, \qquad [X,Y] = 0, 
$$
satisfying the condition of vanishing twist constants \eqref{vani}.  
We also prescribe a scalar field $\rhob$ and a vector field $\jb$ defined on $T^3$, also 
satisfying the Gowdy symmetry assumption, and we require the regularity 
\be
\label{data}
\gb \in H^1(T^3),  \qquad \kb \in L^2(T^3), 
\qquad 
\rhob, \, \jb  \in L^1(T^3).
\ee
(As already pointed out, we actually cover slightly more general data in which certain components 
of the fields $\gb, \kb$ are less regular.) 
Finally, we assume that these data satisfy Einstein's constraint equations
\be
\label{constraints}
\aligned 
& \Rb + \big( \tr \kb \big)^2 - \big| \kb \big|^2 = 2 \kappa \, \rhob, 
\qquad 
\tr\big(\nablab \, \kb\big) - \nablab \big( \tr \kb \big) = \kappa \jb,
\endaligned 
\ee
where $\nablab$ and $\Rb$ denote the covariant derivative operator and scalar curvature of $(T^3, \gb)$, respectively, 
and the trace and norm are determined from the metric $\gb$. 
Under the regularity \eqref{data}, 
the curvature terms $\Rb, \tr\big( \nablab \, \kb \big)$, and $\nablab \big(\tr \kb\big)$ are
well-defined as distributions~\cite{LeFlochMardare}, while the remaining terms in \eqref{constraints} belong to $L^1(T^3)$. Of course, 
\eqref{constraints} actually implies that the scalar field $\Rb$ is more regular and, actually, integrable in space. 
In fact, we will not use directly the definition from \cite{LeFlochMardare}; instead, 
later in this text, a complete definition of weak solutions appropriate to the problem under consideration 
will be introduced.

We refer to $\big( \gb, \kb,\rhob,\jb \big)$ as a {\sl finite energy, Gowdy symmetric,
initial data set on $T^3$ for the Einstein--Euler equations.}  

We are now in a position to state the main result of the present paper. Recall first that 
an important quantity associated with Gowdy-symmetric spacetimes is the {\sl area function} $R$ 
(defined up to a multiplicative constant) of the orbits of the symmetry group generated by the Killing fields. 
Since the metric coefficients are only in $H^1$, the gradient $\nabla R$ may only be defined
almost everywhere for the Lebesgue measure. However, using the Einstein equations we will prove later
(cf.~Proposition~\ref{prop52}, below)
that $\nabla R$ is actually {\sl continuous}. In addition, 
a standard argument showing that $\nabla R$ is a {\sl timelike} vector field 
remains valid under our regularity condition and even for compressible matter. 
This condition on the area function can be expressed explicitly in terms of the initial data: when the spacetime 
metric is sought for in conformal coordinates (see \eqref{confo}, below), the initial data prescribed 
for the time derivative of the area function
(denoted by $\Rbz$ below in Section~\ref{contra})
should be everywhere positive (expanding case) or everywhere negative (contracting case).

This property allows us to distinguish between two cases, depending whe\-ther  
the spacetime is {\sl expanding} or {\sl contracting,} that is, the area of the orbits of symmetry is 
increasing or decreasing toward the future, respectively. Furthermore, without genuine loss of generality
we assume that the initial hypersurface has constant area function, which imposes certain restrictions
on the initial data set but, experience with (vacuum and regular, at least) Gowdy symmetric spacetimes 
tells us that this is not a restriction on the class of spacetimes themselves, only
a minor restriction on the choice of the initial hypersurface.

The following main theorem encompasses both the expanding and the contracting cases.

\begin{theorem}[Einstein--Euler spacetimes with Gowdy symmetry on $T^3$] 
\label{maintheo} 
Let $\big( \gb, \kb,\rhob,\jb \big)$ be a finite energy, Gowdy symmetric, initial data set on $T^3$
for the Einstein--Euler equations, and assume that these initial data have constant area,  
and are everywhere expanding or 
everywhere contracting. Then, there exists a finite energy, Gowdy symmetric spacetime $(M, g, \rho, j)$
satisfying the Einstein--Euler equations \eqref{Einstein}--\eqref{pressure}
in the distributional sense, and the following properties hold.
The manifold $(M,g,\rho, j)$ is (up to diffeomorphisms) a Gowdy-symmetric 
future development of 
$\big( \gb,\kb,\rhob,\jb \big)$,  
which is globally covered by a single chart of coordinates $t$ and $(\theta,x,y) \in T^3$, with 
$$
M = 
\begin{cases}
\quad \big\{ (t, \theta) \, / \, 0 < c_0 \leq t < \infty \big\} \times T^2,   &\text{ expanding case,} 
\\
\\ 
\quad \big\{ (t, \theta) \, / \, c_0 \leq t < c_1 \leq 0 \big\} \times T^2,   &\text{ contracting case.} 
\end{cases}
$$ 
Here, $c_1 \in (c_0,0]$ is a constant, and the time variable is chosen to coincide  
with the area of the surface of symmetry in the expanding case,
 and with minus this area in the contracting case.  
\end{theorem}  

The above theorem relies on a notion of weak solution which will be described in full detail in this text (in Sections~4 and 5); importantly, 
this definition incorporates the initial data in a weak sense as well. 

It is interesting to ask if $c_1$ can be taken to be
equal to zero. Note for comparison that if the perfect fluid occurring in
the theorem is replaced by collisionless matter described by the Vlasov
equation then the analogous theorem does hold with $c_1=0$, as was established in \cite{weaver}.
For a fluid with a linear equation of state there are solutions for which
the theorem does not hold with $c_1=0$ as follows from the discussion in
Section~12.2 of \cite{rendall08}. At the same time it seems, as will be 
discussed in more detail in Section~\ref{homogeneous}, that these solutions 
are exceptional in this regard.

The above statement provides a global foliation   
of the constructed spacetimes. 
Due to the {\sl time irreversibility} of discontinuous solutions to the Euler equations,
compressible matter spacetimes can only be defined in the {\sl future} 
of the initial hypersurface. This is in contrast with vacuum spacetimes and Vlasov spacetimes 
which can be defined in both time directions.

An important and very challenging open problem is to establish the strong cosmic censorship 
(that is, the inextendibility of the future Cauchy development)
for the spacetimes constructed in Theorem~\ref{maintheo}.  
In the expanding case, the future inextendibility of the spacetime follows from the fact 
that the area function tends to infinity along every future timelike direction. 
The generic inextendibility of (vacuum) Gowdy spacetimes in the contracting direction 
was established in major contributions by  
Chru\'sciel, Isenberg, and Moncrief \cite{ChruscielIsenbergMoncrief}
(polarized case) and Ringstr\"om  \cite{Ringstrom1,Ringstrom2}, 
and recently also established by Dafermos and Rendall for the Vlasov model \cite{DafermosRendall}
(cf.~Theorem 4.1 therein).

Observe that the uniqueness issue is not addressed here, since 
within the functional framework under consideration 
such a result is not even known for the Euler equations in the flat Minkowski geometry.  
The entropy inequalities are however fully motivated, and are known to imply uniqueness for certain
restricted classes of initial value problems or smaller classes of regularity of solutions; 
we refer the reader to \cite{Dafermos-book,LeFloch-book} for details. 
Note also that although the areal foliation constructed in Theorem~\ref{maintheo} 
is geometric in nature, it is tied to the Gowdy-symmetry assumption, and 
it would be interesting to investigate the existence of a global foliation 
by constant mean curvature (CMC) slices.

%==================================================================================================================

\section{Einstein--Euler equations}
\label{EE}

\subsection*{A $(3+1)$-decomposition of the Euler equations}

We present here a formulation of the Euler equations in a divergence form, 
which we will later use to define a notion of weak solutions. All calculations in the present section
are carried out in the smooth setting. 
Recall that the Bianchi identities for the geometry implies the Euler equations for the fluid 
\be
\label{Euler}
\nabla^\alpha T^\beta_\alpha = 0,
\ee
in which the energy-momentum tensor determined by \eqref{momentum}. 
We will eventually express these equations in local coordinates adapted to the Gowdy symmetry
 but, 
as a preliminary step,
we consider a general foliation by spacelike hypersurfaces, determined as 
the level sets 
of a time-function $t$. Choosing a vanishing shift vector and 
introducing the future-oriented unit normal $n^\alpha$,
the lapse function $N>0$, vanishing shift-vector, and the 
second fundamental form $k^{\alpha\beta}$ to the foliation, we can write  
$$
\aligned
& n^\alpha := - N \nabla^\alpha t, 
\qquad
&& N^{-2} := - g(\nabla t, \nabla t), 
\\
& k(X,Y) := g( \nabla_X Y, n) = - g(\nabla_Y n,x).   &&  
\endaligned
$$
Here, for each hypersurface, 
$X,Y$ denote arbitrary vector fields tangent to any given hypersurface (with $Y$ arbitrarily extended to a neighborhood
of that slice). 

Using the normal $n$ we determine the projection operator $h^{\alpha \beta} := g^{\alpha\beta} + n^\alpha n^\beta$, 
satisfying the obvious conditions $h^{\alpha\beta} n_\alpha=0$
and $h^{\alpha\beta}X_\alpha = X^\beta$ whenever $n_\alpha X^\alpha = 0$. 
We decompose the matter tensor $T^{\alpha\beta}$ into its normal and tangential components, as follows: 
\be
\label{rhojS} 
\aligned 
\rho & := T^{\alpha\beta} n_\alpha n_\beta, 
\qquad 
j^\alpha := - T^{\beta\gamma} h^\alpha_\beta n_\gamma, 
\qquad 
S^{\alpha\beta} := T^{\gamma\delta} h^\alpha_\gamma h^\beta_\delta, 
\endaligned 
\ee
so that $T^{\alpha\beta} = \rho n^\alpha n^\beta + j^\alpha n^\beta  + j^\beta n^\alpha+ S^{\alpha \beta}$. 
The scalar $\rho$ is the mass-energy density measured by an observer moving orthogonally to 
the slices of the foliation and $j$ is its momentum vector which is tangent to the hypersurfaces. 
Using the expression \eqref{momentum} of the energy-momentum tensor, we find
$$
\aligned
& \rho = (\mu + p) \big( u_\alpha n^\alpha\big)^2 - p, 
\qquad 
 j^\alpha = - (\mu +p) \big( u_\beta n^\beta\big) u^\gamma h^\alpha_\gamma.  
\endaligned 
$$
Note that $T^{\alpha\beta} u_\alpha u_\beta=\mu$, 
so that $\mu$ is the mass-energy density measured by a (Lagrangian) observer moving with the fluid.  

In coordinates $(t,x^a)$ ($a=1,2,3$) adapted to the foliation so that $t$ is constant on each slice, 
we have $(g_{\alpha\beta}) = -N^2 \, dt^2 + (g_{ab})$ and   
\be
\label{1122}
\aligned 
n_0 &= -N,  \qquad    \qquad  & n_a = 0,        \qquad  &k_{ab} = - {1 \over 2 N} \del_t g_{ab},  
\\
\rho &= N^2 \, T^{00},   \qquad & j^0 = 0,     \qquad             &j^a = N \, T^{0a}, 
\\
S^{00} &= 0,       &S^{0a} = 0,                \qquad & S^{ab}  = T^{cd} h^a_c h^b_d. 
\endaligned 
\ee
Since $u^\alpha n_\alpha = u^0 n_0 = - N \, u^0$ and $-1 = u^\alpha u_\alpha = - N^2 (u^0)^2 + g_{ab} u^a u^b$, 
we thus find 
\be
\label{2233}
\aligned  
\rho & = (\mu+p) N^2 (u^0)^2 - p, 
\qquad
j^a =  (\mu+p) N u^0 u^a, 
\\
S^{ab} & = (\mu+p) \, u^a u^b + p \, g^{ab}. 
\endaligned 
\ee
Now, writing the Euler equations \eqref{Euler} as 
$$
\aligned 
0 = \nabla^\alpha T_\alpha^\beta 
& = \nabla^\alpha\left(\rho \, \big( n_\alpha + j_\alpha \big) n^\beta + j^\beta n_\alpha 
+ S_\alpha^\beta \right) 
\\
& = n^\beta \nabla^\alpha\left(\rho n_\alpha + j_\alpha \right) 
+ \nabla^\alpha S_\alpha^\beta + (\nabla^\alpha j^\beta)  n_\alpha  
    +  \nabla^\alpha n^\beta \left(\rho n_\alpha + j_\alpha \right) 
    + j^\beta \nabla^\alpha n_\alpha, 
\endaligned 
$$
and decomposing them into normal and tangential components, we find 
$$
\aligned 
& 0 = - \nabla^\alpha\left(\rho n_\alpha + j_\alpha \right) + n_\beta \nabla^\alpha S_\alpha^\beta
          + n_\alpha n_\beta (\nabla^\alpha j^\beta), 
\\
& 0 = h_\gamma^\beta \nabla^\alpha j^\gamma n_\alpha + h^\beta_\gamma \nabla^\alpha S_\alpha^\gamma
    +  h^\beta_\gamma \nabla^\alpha n^\gamma \left(\rho n_\alpha + j_\alpha \right)
    +    j^\beta \nabla^\alpha n_\alpha. 
\endaligned  
$$
 
At this juncture, we observe that in order to define a concept of solution to the Einstein--Euler equations
as we will do later, it is important to keep these equations in divergence form.    
The suitable form of the equations consistent with the definition of distributional geometry 
presented in \cite{LeFlochMardare}
will be obtained provided
we take into account the volume element associated with the Riemannian metric $\gthree$
induced on the slices, as well as
the volume element associated with the Lorentzian metric $g$.
With this aim in mind, we introduce the rescaled matter fields defined as follows: 
\be
\label{3344}
\aligned 
& \rhot :=  \omega \, \rho,  \qquad 
&& \omega := (\det \gthree)^{1/2}, 
\\
&\jtn := N \, \jt   := N \,  \omega \, j,  
\qquad 
&& \Stn  := N^2 \, \St  := N^2 \,  \omega \, S. 
\endaligned 
\ee
The scalars $\mu$ and $p$ are rescaled in the same way as $\rho$, with an obvious notation. 
Observe that scalar, vector, and $2$-tensor fields are scaled differently. 

After some further calculations, the Euler equations \eqref{Euler} take the form 
\be
\label{Euler00-prime}
\aligned 
& \del_t \rhot + \del_a \jtn^a = \Sigmat_1,  
\\
& \del_t \jtn^a + \omega \, {}^{(3)}\nabla_b \big( \omega^{-1} \Stn^{ab} \big) 
 = \Sigmah_2^a,  
\endaligned 
\ee  
where ${}^{(3)}\nabla$ is the connection induced on the hypersurfaces, and 
\be
\label{Euler00-primeBIS}
\aligned
& \Sigmat_1 := N^{-1} \big( - \del_a N \, \jtn^a  + k_{ab} \, \Stn^{ab} \big), 
\\
& \Sigmah_2^a := N^{-1} \big(  \del_t N \, \jtn^a + \del_b N \, \Stn^{ab} \big) 
+ 2N \, {k^a}_b \, \jtn^b - \rhot \, N \nabla^a N. 
\endaligned 
\ee
This completes the derivation of a general $(3+1)$-decomposition of the Euler equations. Note that the equations
\eqref{Euler00-prime} have a divergence form. 

Now, under the Gowdy symmetry assumption  the Euler equations simplify,
provided we use coordinates $x=(t, x^1, x^B)$ (with $B=2,3$) adapted to this symmetry, so that 
the coordinates $(x^2, x^3) \in T^2$ describe the $2$-surfaces spanned by the Killing fields $K,L$.  
In agreement with the derivation of \eqref{Euler00-prime}-\eqref{Euler00-primeBIS}, 
our choice of coordinates $(t, x^1)$ is made so that $g_{01}=0$ 
(see the metric expression \eqref{are} and \eqref{confo}, below). 
Furthermore, imposing 
\be
\label{797} 
u^2=u^3=0
\ee
 is compatible with our symmetry assumptions. 
So, recalling that $u^\alpha$ is a future-oriented, time-like, unit vector, we find   
\be
\label{velocity} 
N^2 (u^0)^2 - g_{11} (u^1)^2 = 1. 
\ee 
The velocity vector is determined by the single function $u^0$.

The (non-vanishing) components of the matter tensor then take the form (with $B,C=2,3$) 
$$
\aligned
\rho & = (\mu + p) N^2 (u^0)^2 - p, 
\qquad
j^1 = (\mu+p) N u^0 u^1, 
\\
%S^1 := 
S^{11} & = (\mu+p) \, (u^1)^2 + p \, g^{11}, 
\qquad S^{1B} = 0, \qquad S^{BC} = p \, g^{BC}. 
\endaligned 
$$ 
From their definition, we easily obtain the Christoffel symbols 
$$
\aligned 
& \Gamma^1_{BC} = - {1 \over 2} g^{11} \del_1 g_{BC}, 
\qquad 
\Gamma_{11}^1 = {1 \over 2} g^{11} \, \del_1 g_{11},
\endaligned 
$$ 
and we are in a position to compute explicitly a key term in \eqref{Euler00-prime}: 
$$
\aligned 
\omega \, {}^{(3)} \nabla_b \big( \omega^{-1} \Stn^{1b} \big) 
& = \del_1 \Stn^{11} + \Gamma^b_{bc} \Stn^{1c} + \Gamma^1_{bc} \Stn^{bc} - \omega^{-1} \nabla_b \omega S^{1b} 
\\
& = \del_1 \Stn^{11} + \Gamma^1_{11} \Stn^{11} + \Gamma^1_{BC} \Stn^{BC}. 
\endaligned 
$$

Recalling that $\mut$ and $\pt$ are defined by rescaling as $\rhot$ in \eqref{3344}, 
we arrive at the formulation 
\be
\label{Euler00-prime22}
\aligned 
& \del_t \rhot + \del_1 \jtn^1 = \Sigmat_1,  
\\
& \del_t \jtn^1 + \del_1 \Stn^{11} = \Sigmah_2,  
\endaligned 
\ee  
the right-hand terms being given by  
\be
\label{Euler11}
\aligned 
\Sigmat_1
:= & N \, \big( k_{11} \, (\mut + \pt) \, (u^1)^2 + (\tr k) \, \pt \big) - N \, \del_1 N \, (\mut + \pt) \, u^0 u^1, 
\\
\Sigmah_2 = \Sigmah_2^1 
:= \, &  N \, \left( \del_t N \, (\mut + \pt) \, u^0 u^1  + \del_1 N \, \big( (\mut + \pt) \, (u^1)^2  + \pt \, g^{11} \big) \right)  
\\
& - {N^2 \over 2} \, \Big(  g^{11} \del_1 g_{11}  \big( (\mut + \pt) (u^1)^2 + \pt \, g^{11} \big)
  -  g^{11} g^{BC} \del_1 g_{BC} \pt 
\Big)
\\
& + 2 \, N^3 k^1_1 \, ( \mut + \pt) \, u^0 u^1 - \rhot \, g^{11} \,  \del_1 N.
%%%%%%%%%%%%%%%%                                          - \rhot \, N g^{11} \, N \del_1 N.
\endaligned 
\ee

Denoting by $\gtwo$ the determinant of the surface of symmetry and observing that 
\be
\label{4455}
\aligned 
& g^{BC} \del_1 g_{BC} = {1 \over \det (\gtwo)} \del_1 ( \det \gtwo ),
\\
& \tr \, k = -{1 \over 2 N} \, {1 \over \det (\gthree)} \del_t ( \det \gthree ) 
           = - {1 \over 2N} \, \del_t (\ln \omega), 
\endaligned 
\ee
we conclude that in order to express the right-hand sides $\Sigma_1, \Sigma_2$ of the Euler equations \eqref{Euler00-prime22}, 
we need compute only the zero-order terms
$$
N, \quad g^{11}, \quad g_{11}, \quad \det \gthree, \quad \det \gtwo, 
$$
and the first-order terms 
$$
\del_t N, \quad \del_1 N , \quad \del_1 g_{11}, \quad \del_1 ( \det \gtwo), 
\quad \del_1 (\det \gthree), \quad k^1_1. 
$$
Based on \eqref{Euler00-prime22}--\eqref{4455}, we are now in a position to write the field equations in local coordinates.

%----------------------------------------------------------------------------------------------------

\subsection*{Areal coordinates} 
\label{section32}

Recall that this discussion is carried out in the smooth class; weak solutions in areal coordinates
will be discussed in Section~4, below. 
We begin with the so-called areal coordinates $(t,\theta,x^2,x^3)$ in which 
the metric is determined by four scalar functions $U, A, \eta, \alpha$, 
depending on the variables $(t,\theta)$ only: 
\be
\label{are}
g = e^{2(\eta-U)} \, (- \alpha \, dt^2 + d\theta^2) + e^{2U} (dx^2 + A \, dx^3)^2 + e^{-2U} \, t^2 \, (dx^3)^2
\ee
and the variables $\theta,x^2,x^3$ range over the interval $[0,2\pi]$ and are $2\pi$-periodic.  
The coordinates $t, \theta$ parameterize the quotient manifold $M/T^2$, while 
$\del/\del x^2$ and $\del/ \del x^3$ denote the Killing fields and 
$x^2, x^3$ are coordinates on the torus $T^2$. By construction, the area of the 
two-dimensional spacelike orbits of symmetry coincides with the time variable, since 
$$
\mbox{det} \left( 
\begin{matrix}
e^{2U} & A \,  e^{2U}
\\
A \,  e^{2U} & e^{-2U} \, t^2 + A^2 \, e^{2U}
\end{matrix}\right)
= t^2.
$$ 
The fields of $1$-forms  
$$
e_0 := \alpha^{1/2} e^{(\eta-U)} \, dt, \quad 
e_1 := e^{\eta-U} d\theta, \quad 
e_2 := e^U \, ( dx^2 + A \, dx^3), \quad 
e_3 := e^{-U} \, t \, dx^3 
$$
determine an orthonormal frame, whose associated dual frame is  
$$
\alpha^{-1/2} e^{-(\eta-U)} \, {\del \over \del t}, \quad 
e^{-(\eta-U)} {\del \over \del \theta}, 
\quad 
e^{-U} {\del \over \del x^2}, \qquad 
e^U {1 \over t} \, \Big( - A \, {\del \over \del x^2} + {\del \over \del x^3}\Big).  
$$

Returning to the matter variables and recalling the ``projected'' energy-momentum tensor $S^{ab}$ introduced in \eqref{rhojS}, 
we extract its essential components  
\be
\label{PPS}
\aligned 
P_1 & := S(e_1,e_1) = e^{2\eta-2U} \, S^{1}, 
\\
P_2 & := S(e_2,e_2) = e^{2U} \, \big( T^{22} + 2 \, A \, T^{23} + A^2 \, T^{33} \big), 
\\
P_3 & := S(e_3,e_3) = e^{-2U} \, t^2 \, T^{33}, 
\qquad 
P_{23} := S(e_2,e_3) = t \, T^{23} + t A \, T^{33}. 
\endaligned 
\ee  

From now on, we use subscript indices to denote partial derivatives with respect to $t, \theta$.  
After a tedious calculation,  
Einstein's evolution equations extracted from \eqref{Einstein}
take the form of {\sl three nonlinear wave equations}
for the metric coefficients $U,A,\eta$
\be
\label{1}
\aligned 
U_{tt} - \alpha \, U_{\theta\theta} 
& = \Big(- {1 \over t} + {\alpha_t \over 2\alpha} \Big) \, U_t + {\alpha_\theta \over 2} \, U_\theta   
+
\Omega^U(\alpha,t) + \alpha \, \Pi^U(t), 
\\ 
A_{tt} - \alpha \, A_{\theta\theta} 
& = \Big( {1 \over t} + {\alpha_t \over 2\alpha} \Big) \, A_t + {\alpha_\theta \over 2} \, A_\theta  
+ \Omega^A(\alpha) + \alpha \, t \, \Pi^A, 
\\
\eta_{tt} - \alpha \, \eta_{\theta\theta} 
& =  {\alpha_t \over 2\alpha} \, \eta_t  + {\alpha_\theta \over 2} \, \eta_\theta  
    + {1 \over 2} \, \alpha_{\theta\theta} - {\alpha_\theta^2 \over 4 \alpha} 
    + \Omega^\eta(\alpha,t) + \alpha \, \Pi^\eta(t), 
\endaligned 
\ee
in which the lower-order metric terms are  
$$
\aligned 
\Omega^U (\alpha,t) & :=
            {e^{4U} \over 2 t^2} (A_t^2 - \alpha \, A_\theta^2), 
\qquad
\Omega^A(\alpha)  := - 4 (U_t A_t - \alpha \, U_\theta A_\theta), 
\\
\Omega^\eta(\alpha,t) & := (-U_t^2 + \alpha \, U_\theta^2) 
  + {e^{4U} \over 4 t^2} (A_t^2 - \alpha \, A_\theta^2), 
\endaligned 
$$ 
and the lower-order matter terms are  
$$
\aligned 
\Pi^U & := {1 \over 2} e^{2(\eta-U)} \, (\rho - P_1+P_2-P_3), 
&&
\Pi^A := 2 \, e^{2(\eta-2U)} \, P_{23}, 
\\ 
\Pi^\eta(t) & := - {1 \over t^2} A^2 \, e^{2(\eta+U)} P_2 
           - e^{2(\eta-U)} P_3 - {2 \over t} A \, e^{2\eta} P_{23}. 
           &&
\endaligned 
$$
On the other hand, Einstein's constraint equations extracted from \eqref{Einstein} provide us with 
{\sl three first-order differential equations}
for the metric unknowns $\eta$ and $\alpha$, that is,   
\be
\label{4}
\aligned 
{1 \over t} \eta_t & = U_t^2 + \alpha \, U_\theta^2 
        + {e^{4U} \over 4t^2} \, (A_t^2 + \alpha \, A_\theta^2) 
        + e^{2(\eta-U)} \alpha \, \rho, 
\\
{1 \over t} \eta_\theta & 
= - {1 \over 2t\alpha} \alpha_\theta 
  +  2 \, U_t U_\theta + {e^{4U} \over 2 t^2} \, A_t A_\theta - e^{3(\eta-U)} \alpha^{1/2} j^1, 
\\
{1 \over t} \alpha_t & = - 2 \, \alpha^2 \, e^{2(\eta-U)} (\rho - P_1 ). 
\endaligned 
\ee

To finally express the Euler equations, we need the zero-order metric quantities: 
$$
\aligned 
& g_{00} = -N^2 = - \alpha \, e^{2(\eta- U)}, \qquad 
& g_{11}  = e^{2(\eta- U)},  \qquad 
& g^{11}  = e^{-2(\eta- U)}, 
\\
& N  = \alpha^{1/2} \, e^{\eta- U},  \qquad 
& \det(\gtwo)  = t^2,   \qquad 
& \det (\gthree) = t^2 \, e^{2(\eta- U)},
\endaligned 
$$ 
as well as first-order ones:  
$$
\aligned 
\tr \, k 
& = - \alpha^{-1/2} \, e^{-(\eta- U)} \big( {1 \over t} + \eta_t - U_t \big),
\\
k_{11} & = - \alpha^{-1/2} e^{\eta - U} (\eta_t - U_t), 
\\
k_1^1 & = g^{11} k_{11} = - \alpha^{-1/2} e^{-(\eta - U)} (\eta_t - U_t). 
\endaligned 
$$
These expressions allow us to rewrite the equations \eqref{Euler11} in areal coordinates:
\be
\label{boiteE}
\aligned 
(\rhot)_t + (\jtn^1)_\theta  
= & 
\Sigma_1, 
\\
(\jtn^1)_t + (\Stn^{11})_\theta = &  
\Sigma_2,
\endaligned 
\ee
with now 
$$
\aligned 
\Sigma_1 =
& - e^{2(\eta-U)}  (\eta_t - U_t) (\mut + \pt) \, (u^1)^2
  -  \Big( {1 \over t} + \eta_t - U_t \Big) \pt
\\
&  - {1 \over 2} \del_\theta \Big( \alpha e^{2(\eta-U)}\Big) ( \mut + \pt) \, u^0 u^1 
\endaligned 
$$ 
and
$$
\aligned 
\Sigma_2 = 
&  {1 \over 2} \, \del_t \big( \alpha \, e^{2(\eta-U)} \big)  \, (\mut + \pt) \, u^0 u^1  
      + {1 \over 2} \, \del_\theta \big( \alpha \, e^{2(\eta-U)} \big) \, 
         \big( (\mut + \pt) \, (u^1)^2  + \pt e^{-2(\eta-U)} \big) 
\\
& - \alpha \, (\eta_\theta - U_\theta) \, \Big(
e^{2(\eta-U)} \, (\mut + \pt) (u^1)^2 + \pt \Big)
\\
& - 2 \alpha e^{2(\eta-U)} \big( \eta_t - U_t \big) \, ( \mut + \pt) \, u^0 u^1 
- {1 \over 2} \rhot \, \big( \alpha_\theta + 2 \alpha (\eta_\theta - U_\theta) \big). 
\endaligned 
$$ 

Finally, to ``close'' the system it is necessary to have explicit expressions of the conservative and flux variables 
$\rhot, \jtn^1, \Stn^{11}$ (arising in \eqref{boiteE})
in terms of the energy density $\mu$ and the velocity function 
$$
\vv := \alpha^{-1/2} \, {u^1 \over u^0}. 
$$
Noting that $\alpha \, (u^0)^2 - (u^1)^2 = e^{-2(\eta- U)}$, we obtain 
$$
e^{2(\eta- U)} \, (u^0)^2 = {1 \over \alpha(1 - \vv^2)}, 
\qquad 
e^{2(\eta- U)} \, (u^1)^2 = {V^2 \over 1 - \vv^2}. 
$$
In view of $\rho = \mu + (p+\mu) e^{2(\eta- U)} (u^1)^2$, the rescaled mass-energy density reads 
\be
\label{density5}
\rhot =  t \, e^{(\eta- U)} \Big( \mu + (p+\mu) {\vv^2 \over 1 - \vv^2}\Big). 
\ee
For the momentum we find 
$j^1 = e^{-(\eta- U)} \, (\mu +p)  \vv (1 - \vv^2)^{-1}$, 
so that the rescaled momentum variable reads 
\be
\label{momentum5}
\jtn^1 = t \,  \alpha^{1/2} e^{\eta- U} \, (\mu +p) \, {\vv \over 1 - \vv^2}. 
\ee 
Finally, from $\Stn^{11} = t \alpha e^{3(\eta-U)} \Big(  (\mu + p) (u^1)^2 + p g^{11} \Big)$ we obtain  
\be
\label{stress}
\Stn^{11} = t e^{(\eta-U)} \alpha \, \Big( (\mu + p) {\vv^2 \over 1 - \vv^2} + p \Big).  
\ee

Consequently, we can express $\Sigma_1$ and $\Sigma_2$ in terms of $\mu, V$ and obtain 
$$
\aligned 
\Sigma_1  
= & - e^{\eta - U} \left( t \, ( \eta_t - U_t) (\mu + p) \, {\vv^2 \over 1 - \vv^2}
                 + ( 1 + t (\eta_t - U_t) ) \, p \right) 
\\
&  - t  \alpha^{1/2} e^{\eta - U} \big( {\alpha_\theta \over 2\alpha } + \eta_\theta - U_\theta \big) ( \mu + p) \,  {\vv \over 1 - \vv^2} 
\endaligned 
$$ 
and
$$
\aligned 
\Sigma_2 
 = 
& {t \over 2} \, \alpha^{-1/2} e^{\eta-U} \, \big( \alpha_t + 2 \alpha (\eta_t - U_t) \big)  
     \, (\mu + p) \, {V \over 1-V^2}  
\\
& + {t \over 2} \, e^{\eta-U} \, \big( \alpha_\theta + 2 \alpha (\eta_\theta - U_\theta) \big)  
         \big( (\mu + p) \, {V^2 \over 1-V^2}  + p \big) 
\\
& - 2t \, \alpha \, (\eta_\theta - U_\theta) \, e^{\eta-U} \, \Big( (\mu + p) {V^2 \over 1-V^2}  + p \Big)
\\
& - 2 t \, \alpha^{1/2} e^{\eta-U} \big( \eta_t - U_t \big) \, ( \mu + p) \, {V \over 1-V^2},
\endaligned 
$$ 
which completes the derivation of the {\sl Euler equations in areal coordinates}  
\be
\label{Euler-areal0} 
\aligned 
\Bigg(   t \, e^{(\eta- U)} \Big( \mu + (\mu+p) {\vv^2 \over 1 - \vv^2}\Big)   \Bigg)_t 
  + \Bigg( t \, e^{\eta- U} \alpha^{1/2} \, (\mu +p) \, {\vv \over 1 - \vv^2} \Bigg)_\theta  
= \, & \Sigmat_1, 
\\
\Bigg( t \, e^{\eta- U} \alpha^{1/2} \, (\mu +p) \, {\vv \over 1 - \vv^2} \Bigg)_t
+ \Bigg(   t e^{(\eta-U)} \alpha \Big( (\mu + p) {\vv^2 \over 1 - \vv^2} + p \Big) \Bigg)_\theta 
 = \, & \Sigmah_2. 
\endaligned 
\ee

Finally, assuming the linear equation of state \eqref{pressure} we conclude that 
the fluid variables $\mu, V$ satisfy the {\sl nonlinear hyperbolic system of two balance laws}  
\be
\label{Euler-areal2First} 
\aligned 
\Bigg(   t \, e^{\eta- U} {1 + k^2 \, \vv^2 \over 1 - \vv^2} \mu \Bigg)_t 
  + \Bigg( t \, e^{\eta- U} \, \alpha^{1/2} { (1+k^2) \vv \over 1 - \vv^2} \mu \Bigg)_\theta  
= \, & t \, \mu e^{\eta - U}  \Sigma_1'', 
\\
\Bigg( t \, e^{\eta- U} \alpha^{1/2} \, {(1+k^2) \vv \over 1 - \vv^2} \mu \Bigg)_t 
+ \Bigg(   t \, e^{\eta-U} \alpha \, {k^2 + \vv^2 \over 1 - \vv^2} \mu \Bigg)_\theta 
 = \, &  t \, \mu e^{\eta-U}  \Sigma_2'', 
\endaligned 
\ee
with coefficients depending upon the metric functions $U, \eta, \alpha$, and 
$$
\aligned 
\Sigma_1'' 
:= & - \left( {k^2 \over t} + ( \eta_t - U_t)\, {k^2 + \vv^2 \over 1 - \vv^2} \right)  
   - \alpha^{1/2} \left( {\alpha_\theta \over 2\alpha } + \eta_\theta - U_\theta \right) {(1+k^2)\vv \over 1 - \vv^2}
\endaligned 
$$ 
and
$$
\aligned 
\Sigma_2''  
= & \alpha^{-1/2} \, \big( {\alpha_t \over 2} - \alpha (\eta_t - U_t) \big)  
    {(k^2+1) \, V \over 1-V^2}  
  +  \big( {\alpha_\theta \over 2} - \alpha (\eta_\theta - U_\theta) \big)  
         \, {V^2 + k^2 \over 1-V^2}.
\endaligned 
$$  
As we will see later the second equation in \eqref{Euler-areal2First} should still be 
multiplied by $\alpha^{-1}$, leading to the final form of the Euler system
\be
\label{Euler-areal2} 
\aligned 
\Bigg(   t \, e^{\eta- U} {1 + k^2 \, \vv^2 \over 1 - \vv^2} \mu \Bigg)_t 
  + \Bigg( t \, e^{\eta- U} \, \alpha^{1/2} { (1+k^2) \vv \over 1 - \vv^2} \mu \Bigg)_\theta  
= \, & t \, e^{\eta-U} \, {S_1 \over 1 - \vv^2} \, \mu, 
\\
\Bigg( t \, e^{\eta- U} \,  \alpha^{-1/2} {(1+k^2) \vv \over 1 - \vv^2} \mu \Bigg)_t 
+ \Bigg(   t \, e^{\eta-U} \, {k^2 + \vv^2 \over 1 - \vv^2} \mu \Bigg)_\theta 
 = \, &  t \, e^{\eta-U} \, {S_2 \over 1 - \vv^2} \, \mu, 
\endaligned 
\ee
with 
$$
\aligned 
  - S_1 
:= & {k^2 (1-V^2) \over t} + ( \eta_t - U_t)\, (k^2 + \vv^2)  
     + \alpha^{1/2} \left( {\alpha_\theta \over 2\alpha } + \eta_\theta - U_\theta \right) (1+k^2)\vv
\endaligned 
$$ 
and
$$
\aligned 
 - S_2  
: = & \alpha^{-1/2} \, \big( {\alpha_t \over 2 \alpha} + \eta_t - U_t \big)  
    (k^2+1) \, V 
  +  \big( {\alpha_\theta \over 2 \alpha} + \eta_\theta - U_\theta \big)  
         \, (V^2 + k^2).
\endaligned 
$$ 
Observe that the metric coefficient $A$ does not arise in the Euler equations.

It remains to compute now the expressions of $P_1, P_2, P_3, P_{23}$ required in the right-hand side of \eqref{1}. 
Computing the coefficients of the inverse of the metric, specifically the coefficients 
$g^{22}= t^{-2} e^{2U} A^2 + e^{-2U}$,  
$g^{23}= -t^{-2} e^{2U} A$,  
and $g^{33}= t^{-2} e^{2U}$,   
we find 
$$
\aligned 
P_1  
       =  {\mu \vv^2 + p \over 1 - \vv^2} = {\vv^2 + k^2 \over 1 - \vv^2} \, \mu, 
\qquad 
P_2 = P_3  = p = k^2 \, \mu, 
\qquad 
P_{23} = 0 
\endaligned 
$$
and, therefore, by recalling that $\rho = (\mu + p V^2)/(1-V^2)$,  
\be
\label{piUAE} 
\aligned 
\Pi^U & := {1 \over 2} e^{2(\eta-U)} \, (\mu - p) = {1 -k^2 \over 2} e^{2(\eta-U)} \, \mu, 
\quad 
&& \Pi^A := 0, 
\\ 
\Pi^\eta(t) & := - e^{2 \eta} \, k^2 \, \Big( e^{-2U} + {A \over t^2} e^{2U} \Big) \, \mu. 
&&
\endaligned 
\ee 
Note also that the expression of the right-hand of the equation for $\alpha_t$ 
\be
\label{eqalpha}
\aligned
{1 \over t} (1/\alpha)_t 
         & = 2 (1-k^2) \, e^{2(\eta-U)} \, \mu
         \\
         & = 2 {(1-k^2)(1-V^2) \over 1 - k^2 V^2} \, e^{2(\eta-U)} \, \rho. 
\endaligned
\ee

%----------------------------------------------------------------------------------------------------

\subsection*{Conformal coordinates}

Recall that this discussion is carried out in the smooth class; weak solutions in conformal coordinates
will be discussed in Section~5, below. 
One may also use coordinates $(\tau,\theta,x,y)$ 
in which the metric has the ``isothermal'' form
\be
\label{confo}
g = e^{2(\eta-U)} \, (- d\tau^2 + d\theta^2) + e^{2U} (dx + A \, dy)^2 + e^{-2U} \, R^2 \, dy^2 
\ee
and thus depends on the four scalar functions $\eta, U, A$, and $R$ of the variables $(\tau,\theta)$, only.
The area $R$ of the surfaces of symmetry is now an independent unknown function.  
Formally, the function $\alpha$ is replaced by $1$, and the variable $t$ is replaced by $R$, and one therefore still has 
to determine four metric coefficients.  

In comparison with the areal coordinates, the Einstein--Euler equations take a somewhat simpler form in conformal coordinates. There are now four evolution equations: 
\be
\label{11}
\aligned 
U_{\tau\tau} - U_{\theta\theta} 
& = \Omegat^U + \Pit^U, 
\\ 
A_{\tau\tau} - A_{\theta\theta} 
& = \Omegat^A(1) + R \, \Pit^A, 
\\
\eta_{\tau\tau} - \eta_{\theta\theta} 
& = \Omegat^\eta + \Pit^\eta, 
\\
R_{\tau\tau} - R_{\theta\theta} & = \Pit^R, 
\endaligned 
\ee
where 
$$
\aligned 
& \Omegat^U := - {1 \over R} (R_\tau U_\tau - R_\theta U_\theta) + \Omega^U(1,R),
\\
& \Omegat^A := {1 \over R} (R_\tau A_\tau - R_\theta A_\theta) + \Omega^A(1), 
\quad 
&& \Omegat^\eta := \Omega^\eta(1,R), 
\endaligned
$$
and 
$$
\aligned 
& \Pit^U := \Pi^U(R), \qquad \Pit^A := \Pi^A, \qquad \Pit^\eta := \Pi^\eta(R),
\\
& \Pit^R := R e^{2(\tau-U)} (\rho - P_1). 
\endaligned 
$$
Hence, the lower-order terms are given by (essentially) the same expressions as in Section~\ref{section32},
the $t$-derivatives being replaced by $\tau$-derivatives, and 
$\alpha, t$ replaced by $1, R$, respectively. 
In conformal coordinates, there are only {\sl two constraint equations}
\be
\label{4-conf}
\aligned 
{R_{\theta\theta} \over R}  
& = {1 \over R} (\eta_\tau R_\tau + \eta_\theta R_\theta)  
     - (U_\tau^2 + U_\theta^2) 
     - {e^{4U} \over 4R^2} \, (A_\tau^2 + A_\theta^2) - e^{2(\eta-U)} \, \rho, 
\\
{R_{\tau\theta} \over R}  
& = {1 \over R} (\eta_\tau R_\theta + \eta_\theta R_\tau)  
   - 2 \, U_\tau U_\theta - {e^{4U} \over 2 R^2} \, A_\tau A_\theta + e^{2(\eta-U)} j^1,
\endaligned 
\ee
which can be regarded as {\sl first-order differential equations} for $R_\theta$. 
Furthermore, the Euler equations \eqref{Euler-areal2}  can similarly be written in conformal coordinates.

%------------------------------------------------------------------------------------

\subsection*{Spatially homogeneous solutions}\label{homogeneous}

In this section the equations will be expressed in areal coordinates. A special
type of solutions are those for which all variables are independent of the 
coordinate $\theta$ and so all the evolution equations reduce to ordinary 
differential equations. These are what are known as spatially homogeneous solutions. They
admit an action of a three-dimensional Abelian group by symmetries and so in
the usual terminology of general relativity they are of Bianchi type I. It 
follows from the momentum constraint that $u^1=0$. By a linear transformation 
of the spatial coordinates the initial data $A(0)$ and $\del_t A(0)$ for the variable 
$A$ can be set to zero. It then follows from the evolution equation for this
quantity that $A$ vanishes at all times. The transformation is naturally 
defined on the universal covering of $M$ and need not project to $M$ itself. 
This fact is, however, irrelevant for the study of the dynamics. Thus in the 
homogeneous case it may be assumed without loss of generality that $A$ vanishes 
identically so that the metric is diagonal. The Bianchi type I solutions of 
the Einstein equations can be 
determined explicitly provided a new time coordinate $T$ is introduced which 
is a function of the areal time coordinate $t$. The explicit form of the
metric is (see for instance \cite{wainwright}, p.~199)
\be
\label{bianchi1}
-a^{2k^2}dT^2+T^{2p_1}a^{2q_1}d\theta^2+T^{2p_2}a^{2q_2}dx^2+T^{2p_3}a^{2q_3}dy^2, 
\ee 
where the $p_i$ and $q_i$ are constants for $i=1,2,3$ which satisfy the Kasner 
relations $p_1+p_2+p_3=1$, \, $p_1^2+p_2^2+p_3^2=1$, and $q_i=\frac23-p_i$. The
function $a(T)$ is defined by $a^{1-k^2}=\alpha+m^2T^{1-k^2}$ where $\alpha$ 
and $m$ are positive constants. Comparing this with the general form of the
metric in areal coordinates reveals that 
$$
t=T^{-p_1+1} \, \big( \alpha+m^2T^{1-k^2}\big)^{\frac{p_1-2/3}{1-w}}. 
$$
This can not be solved explicitly so as to get $T$ as a function
of $t$ and thus an explicit expression for (\ref{bianchi1}) in areal 
coordinates. It is nevertheless clear that the mapping from $T$ to $t$ can 
be inverted uniquely. The initial singularity occurs at $T=0$. The parameter 
$p_1$ belongs to the interval $[-1/3,1]$. Provided $p_1<1$ it
can seen that $t$ goes to zero as $T\to 0$ and this corresponds to the 
situation $c_1=0$ in the theorem. 

An exception occurs when $p_1=1$ since
in that case $c_1$ is strictly positive. The behavior of the geometry as
$t\to c_1$ in this case has been analysed in \cite{rendall08}. The
Kretschmann scalar $R_{\alpha\beta\gamma\delta}R^{\alpha\beta\gamma\delta}$ tends 
uniformly to infinity so that this is really a curvature singularity. On the 
other hand the metric can be extended to $t=c_1$ in such a way that it remains 
continuous and non-degenerate. Moreover, with repect to this extended
metric the hypersurface $t=c_1$ is null. This is what is known as a weak
null singularity. Note that the above calculations show that within the class
of spatially homogeneous solutions the exceptional solutions are nowhere dense. It
is not clear what will happen with more general inhomogeneous solutions but
all available information indicates that further solutions of this kind will
be at best rare and possibly not exist at all.

%==================================================================================================================

\section{Expanding spacetimes}  
\label{Expa}

\subsection*{A weak formulation of the Einstein equations} 

We now discuss weak solutions to the Einstein--Euler equations, by relying on areal coordinates
and considering the case of expanding spacetimes, in which the area function $R$ is increasing toward the future
from an initial value denoted by $c_0$.   
Hence, the initial value problem is posed with data prescribed on the hypersurface $t:=c_0>0$. 
Instead of $\eta$, it will be convenient to use the new variable
$$
\nu := \eta + {1 \over 2} \ln \alpha, 
$$
so that the geometry is now determined by the four scalar functions $U,A, \alpha, \nu$. 
The natural regularity imposed on these functions is determined in the following lemma.

\begin{lemma}[Energy estimate]  
\label{energybound}
In areal coordinates the total energy 
$$
\aligned 
& E=E(t) := \int_{S^1} \Ecal \, \alpha^{-1/2} \, d\theta,
\\
& \Ecal := ( U_t^2 + \alpha \, U_\theta^2)  
             + {1 \over 4t^2} \, e^{4U} \, ( A_t^2 + \alpha \, A_\theta^2) + e^{2(\nu - U)} \, \rho,  
\endaligned 
$$
is a monotone decreasing function of the variable $t \geq c_0$ and, more precisely, 
satisfies  
$$
\aligned 
& {d \over dt} E = - {2 \over t} \int_{S^1} \Dcal \, d\theta,
\\
& \Dcal := \alpha^{-1/2} \, U_t^2 + {1 \over t^2} \, e^{4U} \alpha^{1/2} A_\theta^2 
         + {1 \over 2} \alpha^{-1/2} e^{2(\nu-U)} (\rho + P_3) \, \geq 0. 
\endaligned 
$$
\end{lemma}

The underlying structure is analogous to two quasilinear wave equations for the variables $U, A$, taken to 
have finite total energy 
$$
\int_{S^1} \Big( ( \alpha^{-1/2} U_t^2 + \alpha^{1/2}  \, U_\theta^2)  
             + {1 \over 4t^2} \, e^{4U} \, ( \alpha^{-1/2} A_t^2 + \alpha^{1/2} \, A_\theta^2) \Big) \, d\theta,
$$
coupled to the ``first'' Euler equation for $\rho$ taken to have finite total mass-energy 
$$
\int_{S^1}  \rho \,  e^{2(\nu - U)} \, \alpha^{-1/2} \, d\theta. 
$$
Even in the class of weak solutions, the relations in the above lemma hold as {\sl equalities} in the sense of distributions.

In view of the expression of $\Ecal$, as long as 
$\alpha, \alpha^{-1}, U, \nu$ remain bounded, 
i.e.   
\be
\label{boundalpha}
|\alpha| + |\alpha|^{-1} + |U| + |\nu| \lesssim 1, 
\ee 
and provided the initial energy $E(c_0)$ is finite, which we now assume on every compact subset of $[c_0, +\infty)$, 
 Lemma~\ref{energybound} allows us to control 
the $L^2$ norm of the functions $U_t,U_\theta, A_t, A_\theta$ on every slice of the foliation,
as well as the $L^1$ norm of the mass density, from the same quantities evaluated on the initial slice. 
Recalling the expression \eqref{density5} of the density in terms of the fluid variables $\mu, V$, 
we see that the energy estimate imposes 
\be
\label{regul1}
U_t, U_\theta, A_t, A_\theta \in L^\infty_t(L_\theta^2), \qquad 
\muh := {\mu \over 1-V^2} \in L^\infty_t(L_\theta^1),  
\ee
at least on every compact set in time.
In view of \eqref{1}, we see that $U$ and $A$ satisfy
(second-order) wave equations and that two initial conditions must be prescribed for each variable, 
that is, 
$$
\aligned
& U_t(c_0) = \Ubz \in L^2_\theta, \qquad U(c_0)=\Ub \in H^1_\theta, 
\\
& A_t(c_0)= \Abz \in L^2_\theta, \qquad A(c_0) = \Ab \in H^1_\theta. 
\endaligned
$$
The initial data for $\nu$ is then computed from the constraint equation \eqref{4}; see \eqref{500} below. 

Next, in view of the constraint equation (see \eqref{4}) 
\be
\label{NUtheta}
{1 \over t} \nu_\theta 
=  2 \, U_t U_\theta + {1 \over 2 t^2} e^{4U} \, A_t A_\theta - e^{2(\nu-U)} \alpha^{-1/2} j^1, 
\ee
and since $|j^1| \lesssim \rho$ we deduce that 
$\nu_\theta \in L^1_\theta$. Similarly, for the time derivative we have 
\be
\label{NUt}
{1 \over t} \nu_t
= U_t^2 + \alpha \, U_\theta^2 + {e^{4U} \over 4t^2} (A_t^2 + \alpha A_\theta^2) 
+ e^{2(\nu-U)} P_1
\geq 0,   
\ee 
and using $|P_1| \lesssim \rho \in L^1(\Sigma_t)$ we obtain 
$\nu_t \in L^1(\Sigma_t)$.   
It is natural to assume the regularity 
\be
\label{regul2}
\nu_t, \nu_\theta \in L^\infty_t(L^1_\theta).  
\ee
Instead of these differential constraints, we can also use the evolution equation for $\nu$ which follows 
from \eqref{1}, in which case two initial conditions are required on the variable $\nu$
$$
\nu_t(c_0)= \nubz \in L^1_\theta, \qquad \nu(c_0) = \nub \in W^{1,1}_\theta. 
$$ 

Finally, the constraint equation on the function $\alpha$ (see \eqref{4} again) shows that 
\be
\label{regul3}
{\alpha_t \over 1- V^2} \in L^\infty_t(L^1_\theta),
\ee
while no regularity condition can be imposed on the spatial derivative $\alpha_\theta$. 
One should prescribe $\alpha$ on the initial slice, that is, 
$$
\alpha(c_0) = \alphab \in L^1_\theta. 
$$

Having identified the basic functional spaces of interest for each variable, we now consider the 
algebraic structure of the equations. We must reformulate the Einstein 
equations in a form that makes sense under the limited regularity above, only. 
The following key observation can be checked by direct calculation from \eqref{1}.

\begin{proposition}[Weak form of the Einstein equations]
\label{abc} 
The evolution equations for the functions $U$, $A$, and $\nu$ take the form 
\be
\label{weakform}
\aligned 
\big(t \, \alpha^{-1/2} U_t\big)_t - \big(t \, \alpha^{1/2}\, U_\theta\big)_\theta 
& = {e^{4U} \over 2 t \alpha^{1/2}} (A_t^2 - \alpha \, A_\theta^2) 
 + t \, \alpha^{1/2} \, \Pi^U, 
\\ 
\big(t^{-1} \, \alpha^{-1/2} A_t\big)_t - \big(t^{-1} \, \alpha^{1/2}\, A_\theta\big)_\theta 
& = - 4 t^{-1} \, \alpha^{-1/2} \, (U_t A_t - \alpha \, U_\theta A_\theta)
     + \alpha^{1/2} \, \Pi^A, 
\\
\big( t \, \alpha^{-1/2} \big( \nu + (1/2) \log \alpha \big)_t \big)_t - (t \, \alpha^{1/2}\, \nu_\theta)_\theta 
& = t \, \alpha^{-1/2} \, (-U_t^2 + \alpha \, U_\theta^2) 
\\
& \quad   +  {e^{4U} \over 4 t \alpha^{1/2}} (A_t^2 - \alpha \, A_\theta^2) 
   + \alpha^{1/2} \, \Pi^\nu,   
\endaligned 
\ee
where $\Pi^\nu = \Pi^\eta$. 
\end{proposition}

Finally, we supplement \eqref{weakform} with an equation for the function $\alpha$ 
\be
\label{weak-alpha} 
\alpha(t,\theta) 
= \alphab(\theta) \, \exp\Big(- 2 (1-k^2)  \int_0^t t' \, \Big( e^{2(\nu-U)} M \, (1-V^2) \Big) (t', \theta) \, dt' \Big),
\ee
where $\alphab$ is a prescribed data in $L^\infty_\theta$ satisfying $\alphab>0$. Clearly, since $|P_1| \lesssim \rho$ the 
function $\alpha$ is globally bounded: $0 < \alpha \leq \alphab$.

Importantly, in the proposed setting the Einstein equations \eqref{weakform} now make sense 
under the regularity conditions \eqref{regul1}--\eqref{regul3}. 
For instance, $\alpha^{-1/2} U_t$ is the product of an $L^\infty$ function by an $L^\infty_t(L_\theta^2)$ function, 
and its derivative is defined in the distributional sense.   

\begin{remark}
1. The function $\nu$ may be also determined by the constraint equation, that is, 
\be
\label{weak-nu}
\nu(t,\theta) 
=  \int_0^\theta \Big(
2t \, U_t U_\theta + {1 \over 2 t} e^{4U} \, A_t A_\theta - t (1+k^2) e^{\nu -U} \, M \, V
\Big)(t,\theta') \, d\theta', 
\ee
which is an integral equation for the unknown $\nu$.

2. We emphasize that, instead of the equation for $\nu$, the original 
evolution equation for $\eta$ can not be used to derive a~priori estimates, 
since it would require second-order derivatives of $\alpha$ -- which can not be obtained in general 
for the non-vacuum Einstein equations. This is clear from the equation in $\alpha_t$:
it involves the fluid density which generically contains jump discontinuities and, therefore, is not sufficiently regular
to allow us to control second-order derivatives of $\alpha$.  
\end{remark}

%----------------------------------------------------------------------------------------------------------------

\subsection*{Finite energy solutions} 

Let us further discuss the matter variables and introduce a suitable notion 
 of solutions to the Euler equations.
Recalling that $M= \mu/(1-V^2)$,
 and using the variable $\nu$ instead of $\eta$, the equations \eqref{Euler-areal2}
become  
\be
\label{Euler-areal2FINAL} 
\aligned 
\Bigg(   t \, \alpha^{-1/2} e^{\nu - U} (1 + k^2 \, \vv^2) M \Bigg)_t 
  + \Bigg( t \, e^{\nu - U} \, (1+k^2) \vv \, M \Bigg)_\theta  
= \, & t \, e^{\nu-U} \alpha^{-1/2} \, S_1 \, M, 
\\
\Bigg( t \, e^{\nu - U} \,  \alpha^{-1} (1+k^2) \vv \, M \Bigg)_t 
+ \Bigg(   t \, e^{\nu-U} \, \alpha^{-1/2} (k^2 + \vv^2) \, M \Bigg)_\theta 
 = \, &  t \, e^{\nu-U} \alpha^{-1/2}\, S_2 \, M, 
\endaligned 
\ee
with 
$$
\aligned 
  - S_1 
:= & {k^2 (1-V^2) \over t} 
 - {\alpha_t \over 2 \alpha}  \, (k^2 + \vv^2)  
+ ( \nu_t - U_t)\, (k^2 + \vv^2)  
     + \alpha^{1/2} \, ( \nu_\theta - U_\theta ) (1+k^2)\vv
\endaligned 
$$ 
and
$$
\aligned 
 - S_2  
: = & \alpha^{-1/2} \, \big( \nu_t - U_t \big)  
    (k^2+1) \, V 
  +  ( \nu_\theta - U_\theta )  
         \, (V^2 + k^2).
\endaligned 
$$

The selection of physically admissible solutions to the Euler equations 
will be based on the conservation law of the number of particles
$$
\nabla_\alpha N^\alpha = 0
$$
satisfied by sufficiently {\sl smooth} solutions, 
where the vector field $N^\alpha$ has the form 
$N^\alpha = r \, u^\alpha$
with 
$$
{dr \over r} := {dp \over \mu + p(\mu)} 
              = {k^2 \over k^2+1} \, {d\mu \over \mu}, 
$$
hence 
$$
r = \mu^{k^2/(k^2 +1)}.
$$ 
For simplicity, we have assumed here that a single thermodynamical variable 
suffices to determine the state of the fluid under consideration. According to this approximation,  
for the theory to encompass discontinuous solutions, 
we must relax the above equality and impose that solutions satisfy the inequality 
\be
\label{inequality}
- \nabla_\alpha N^\alpha \leq 0, 
\ee 
which we refer to as the {\sl fundamental entropy inequality.}

More generally, an infinite list of (mathematical) entropies is available. 

\begin{definition}
A vector field $\Fcal$ is called an {\sl mathematical entropy flux}
for the Einstein--Euler equations if every smooth solution to these equations satisfies the balance law 
$$
\nabla_\alpha \Fcal^\alpha = \Gcal, 
$$
where $\Fcal^\alpha$ and $\Gcal$ both contain lower-order terms only, that is, zero- and first-order terms in the metric variables
and zero-order terms in the fluid variables. It is said to be convex if the component 
$\Fcal^0$ is convex in the 
conservative variables of the Euler equations (that is, in the variables $(\rhot, \jt^1)$). 
\end{definition}

Physically admissible solutions should then be characterized by the entropy inequalities 
\be
\label{entr}
\nabla_\alpha \Fcal^\alpha \leq \Gcal
\ee 
understood in the distributional sense, for every convex entropy pair that vanishes on the vacuum. 
As presented in \cite{LeFloch}, 
the family of convex weak entropies can be described by an explicit formula which can be used to 
establish a compactness result for sequences of solutions.

%-------------------------------------------------------------------------------------------------------

We are now in a position to state precisely our notion of solutions. 
 
\begin{definition}[Finite energy solutions] 
\label{finit-areal}

1. A {\rm finite energy initial data set for the Einstein--Euler equations} 
is a family of measurable functions $\displaystyle \Ubz, \Ub, \Abz, \Ab, \nubz, \nub, \alphab$, 
$\Mb, \Vb$ satisfying 
$$
\inf_{S^1} \alphab >0, \qquad  \Mb \geq 0, \qquad | \Vb | \leq 1, 
$$
together with the regularity conditions 
$$
\aligned 
& \Ubz, \Abz, \in L^2(S^1), \qquad  
  \Ub, \Ab \in H^1(S^1), 
\qquad  
\nubz, \Mb \in L^1(S^1), 
\\
&  \nub \in W^{1,1}(S^1),
\qquad \alphab, \Vb \in L^\infty(S^1), 
\endaligned 
$$
and the constraint equation
\be
\label{500}
\nub (\theta) = \int_0^\theta \Big( 2 \, c_0 \, \Ubz \, \Ub_\theta + {1 \over 2 c_0} e^{4\Ub} \, \Abz \, \Ab_\theta 
    - c_0 e^{\nub - \Ub} \, (1+k^2) \Mb \, \Vb 
\Big)(\theta') \, d\theta'.  
\ee

2. A {\rm finite energy solution to the Einstein--Euler equations}  
(in areal coordinates) 
is a family of measurable functions $U,A, \alpha, \nu, M,V$ defined on 
$[c_0, \infty) \times S^1$ and satisfying the following conditions: 
\begin{itemize}
\item The functions $\alpha, M, V$ satisfy (for all $t_1> c_0$) 
$$
\inf_{[c_0,t_1] \times S^1} \alpha >0, \qquad  M \geq 0, \qquad | V | \leq 1. 
$$

\item The geometric coefficients $U,A,\alpha,\nu$ satisfy the regularity conditions  
$$
\aligned 
&   U_t, A_t, \in L^\infty_\loc([c_0, +\infty), L^2(S^1)), 
\qquad 
 U, A \in L^\infty_\loc([c_0, +\infty), H^1(S^1)),  
\\
&  
\nu_t, \nu_\theta \in L^\infty_\loc([c_0, +\infty), L^1(S^1)), 
\qquad 
M \, S_1, M \, S_2 \in L^1_\loc([c_0, +\infty) \times S^1).   
\endaligned 
$$ 
\item The functions $U, A, \nu, \alpha$ satisfy the evolution equations \eqref{weakform} 
and the constraint equation \eqref{weak-alpha} in the distributional sense. 
\item The fluid variables $M, V$ satisfy the entropy inequalities 
$$
\nabla_\alpha \Fcal^\alpha \leq \Gcal
$$
in the distributional sense for all convex weak entropy flux $\Fcal^\alpha$ to the Euler equations. 
\item The prescribed initial data are assumed in the sense of distributions. 
\end{itemize}
\end{definition} 

To be more explicit, let us for instance state the first equation in \eqref{weakform} in the distributional sense: 
\be
\label{300}
\aligned
& \int_{c_0}^{+\infty}\int_0^{2 \pi}
\Big(
\psi_t t \, \alpha^{-1/2} U_t - \psi_\theta t \, \alpha^{1/2}\, U_\theta
+ \psi {e^{4U} \over 2 t \alpha^{1/2}} (A_t^2 - \alpha \, A_\theta^2) 
+ \psi t \, \alpha^{1/2} \, \Pi^U
\Big) \, d\theta dt 
\\
& + 
\int_0^{2 \pi} \psi(c_0, \cdot) \, c_0 \, \alphab^{-1/2} \Ubz  \, d\theta =0 
 \endaligned
\ee
for every smooth, compactly supported and spatially $2\pi$-periodic function $\psi: [c_0, \infty) \times [0, 2\pi] \to  \RR$. Observe that the initial data is taken into account in this statement.

Under the condition in the above definition, we will also use the phrase ``spacetimes with finite energy''.
It is important to observe that, under the regularity assumptions under consideration
all the terms arising in the equations under consideration do make sense, at least as distributions. 
In particular, the right-hand sides of the equations \eqref{Euler-areal2FINAL} belong to $L^1$
on every compact subset.

%===========================================================================================================

\section{Contracting spacetimes}  
 \label{contra} 
 
\subsection*{A weak formulation of the Einstein equations}

We continue our discussion of weak solutions to the Einstein--Euler equations, using 
now conformal coordinates in the case that the spacetime is contracting.  
Recall that, in conformal coordinates, the metric is determined by the four functions $U, A, \eta, R$, 
while the fluid is determined by the functions $M,V$. 

It is convenient to pose the problem on a hypersurface of constant negative $\tau_0$. 

\begin{definition}[Finite energy solutions in conformal coordinates] 
\label{finit} 
Given $\tau_0 <0$, 
an {\rm initial data set with finite energy}  for the Einstein--Euler equations in conformal coordinates
is a set of functions 
$$
\aligned 
& \Ubz, \Abz, \in L^2(S^1),
&&  \Ub, \Ab \in W^{1,2}(S^1), 
&&& \etabz \in L^1(S^1), 
&&&& \etab \in W^{1,1}(S^1),
\\
& \Rbz \in L^\infty(S^1),  
&& \Rb \in W^{1,\infty}(S^1),
&&& \inf_{S^1} \Rb >0 
\endaligned
$$
and 
$$
\aligned
& \Mb \in L^1(S^1), \qquad \Vb \in L^\infty(S^1), 
\qquad \Mb \geq 0,   \quad | \Vb | \leq 1. 
\endaligned 
$$
Given an initial data set as above, a set of functions $U,A,\eta,R, M,V$ defined on some interval 
$[\tau_0, \tau_1] \subset [\tau_0,0)$ 
is called a {\rm finite energy solution} to the Einstein--Euler equations in conformal coordinates
if the following conditions hold: 

1. The functions have the regularity
$$
\aligned 
&   U_\tau, A_\tau \in L^\infty([\tau_0, \tau_1], L^2(S^1)), 
&&  U, A \in L^\infty([\tau_0, \tau_1], W^{1,2}(S^1)), 
\\ 
&  \eta_\tau \in L^\infty([\tau_0, \tau_1], L^1(S^1)),   
&&  \eta \in L^\infty([\tau_0, \tau_1], W^{1,1}(S^1)), 
\\
&  R_\tau \in L^\infty([\tau_0, \tau_1], L^\infty(S^1)),   
&&  R \in L^\infty([\tau_0, \tau_1], W^{1,\infty}(S^1)),
&&& \inf_{[\tau_0,\tau_1] \times S^1} R >0 
\endaligned
$$
and
$$
\aligned 
& M \in L^\infty([\tau_0, \tau_1], L^1(S^1)), 
\qquad V \in L^\infty([\tau_0, \tau_1], L^\infty(S^1)), 
\quad 
\\
& M \geq 0,   \qquad | V | \leq 1, 
\qquad
M \, S_1, \, M \, S_2 \in L^1([\tau_0,\tau_1] \times S^1).     
\endaligned 
$$  

2.~The Einstein equations \eqref{11}, \eqref{4-conf} hold in the distributional sense.  

3.~For every convex weak entropy flux $\Fcal^\alpha$ to the Euler equations, the entropy inequality
$$
\nabla_\alpha \Fcal^\alpha \leq \Gcal
$$
holds in the distributional sense. 

4. The following initial data are assumed in the distributional sense: 
$$
\aligned 
& (A_\tau, A)(\tau_0, \cdot) = (\Abz, \Ab), \quad && (U_\tau, U)(\tau_0, \cdot) = (\Ubz, \Ub), 
\\ 
& (\eta_\tau, \eta)(\tau_0, \cdot) = (\etabz, \etab), \quad && (R_\tau, R)(\tau_0, \cdot) = (\Rbz, \Rb),    
\\
& M(\tau_0, \cdot) = \Mb, \qquad && V(\tau_0, \cdot) = \Vb. 
\endaligned 
$$
\end{definition} 
 
The notion of distributional solution is analogous to the one already 
used in the previous section; see in particular the statement \eqref{300} for a specific statement. 
It is important to observe that our regularity conditions are sufficient (and essentially necessary) to define 
in the distributional sense,  
all of the terms appearing in the Einstein--Euler equations.

%----------------------------------------------------------------------------------------------------------------

\subsection*{Monotonicity property of the area function} 

From now on we consider a finite energy solution in the sense of Definition~\ref{finit} and we investigate its geometric properties. 
In particular, by definition, the area function $R$ remains bounded and bounded away from zero.   
We consider the largest region $D^+_\conf$ of the future Cauchy development of the given initial data set 
which can be covered by a single chart in conformal coordinates.

Specifically, following earlier work on classical solutions, 
we investigate the properties of the gradient $\nabla R$ and, 
by combining the evolution and constraint equations satisfied by $R$,  
we establish that 
second-order derivatives of the area function $R$ belong to $L^\infty_\tau(L^1_\theta)$, 
which, in turn, implies that the gradient $\nabla R$ is {\sl continuous}.

\begin{proposition} 
\label{prop52}
The area function $R$ satisfies the following properties: 
\begin{enumerate}
\item The functions $R_\tau, R_\theta$ are continuous in both variables and  
satisfy one of the following alternatives: 
$$
\aligned 
& \text{Case (a) : } \quad R_\tau < - |R_\theta| \quad \text{ for all } \tau, \theta, 
\\
& \text{Case (b) : } \quad R_\tau > |R_\theta|  \quad \text{ for all } \tau, \theta, 
\\
& \text{Case (c) : } \quad   
\text{$R$ is constant and the spacetime is flat and vacuum.}     
\endaligned 
$$
\item In Cases (a) and (b) the gradient of the function $R$ is always timelike. 
\item In Case (a) the components of the gradient $\nabla R$ are uniformly controlled in $L^\infty$: 
$$
|R_\theta| \leq - R_\tau \leq 2 \sup_{S^1} |\Rb_\tau|. 
$$
\item The second-order derivatives $R_{\theta t}$ and $R_{\theta\theta}$ are uniformly controlled 
in $L^\infty_\tau L^1_\theta$: 
\be
\label{301}
\sup_{\tau \geq \tau_0} \int_{S^1} (|R_{\tau\theta}| + |R_{\theta\theta}|) (\tau, \cdot)| \, d\theta
\lesssim  \int_{S^1} (|R_\tau| + |R_\theta|) (\tau_0, \cdot)| \, d\theta 
          + \int_{S^1} \rho (\tau_0, \cdot) \, d\theta. 
\ee
\end{enumerate}
\end{proposition}

In Case (a) (Case (b), respectively)  the spacetime is contracting (expanding, resp.) 
in the future timelike direction.  Our primary interest in this section is on the case of {\sl contracting} spacetimes; 
expanding spacetimes were handled directly in areal coordinates, as discussed in the previous section.

\begin{proof} {\bf 1.} 
We generalize arguments due to Chru\'sciel \cite{Chrusciel} in the case of smooth vacuum spacetimes. 
Consider the functions $\Rp^\pm := R_\tau \pm R_\theta$. 
We obtain 
$$
\aligned 
& \Rp^\pm_\theta + (\eta_\tau - \eta_\theta) \, \Rp^\pm = F^\pm, 
\\
& F^\pm := R (U_\tau \pm U_\theta)^2 + {e^{4U} \over 4R} (A_\tau \pm A_\theta)^2 + R e^{2(\eta-U)} \big( \rho \mp j^1 \big).  
\endaligned
$$
These equations, by assumption, hold in the sense of distributions. However, let us examine the regularity of the various terms involved, as follows. 
Our regularity assumptions show that $(\eta_\tau - \eta_\theta)$ and 
$F^\pm$ belong to $L^\infty_\tau(L^1_\theta)$, so that from our assumption that 
$\Rp^\pm \in L^\infty_\tau(L^\infty_\theta)$
we deduce the stronger regularity property
$$
\Rp^\pm \in L^\infty_\tau(W^{1,1}_\theta).  
$$
In particular, both $\Rp^\pm$ and, therefore, $R_t, R_\theta$ are {\sl continuous} in $\theta$. This 
observation justifies the following calculations. 

Fix some time $\tau$.  
Take any point $\theta_1 \in S^1$ at which the function $R(\tau, \cdot)$ achieves a local extremum value; then, 
since the function $R_\theta$ is continuous, 
we must have $R_\theta(\tau, \theta_1)=0$ and, therefore, 
$$
(\Rp^- \Rp^+)(\tau, \theta_1) = R_\tau(\tau, \theta_1)^2 \geq 0.
$$ 
The case of equality is studied as follows. 
 
For almost every time $\tau$ and every $\theta_0 \in [0, 2\pi]$, we can write 
$$
{\del \over \del \theta} \Big(
\Rp^\pm(\tau, \theta) \, e^{\int_{\theta_0}^{\theta} (\eta_\tau - \eta_\theta)(\tau, \theta') \, d\theta'}\Big)
= 
F^\pm(\tau, \theta) \, e^{\int_{\theta_0}^{\theta} (\eta_\tau - \eta_\theta)(\tau, \theta) \, d\theta'}
\geq 0,  
$$
where we have observed that, since $|J| \leq \rho$, the functions $F^\pm$ are non-negative. 
It follows that 
$$
\aligned
\Rp^\pm(\tau, \theta_0)  
& \leq
\Rp^\pm(\tau, \theta_0+ 2\pi) \, e^{\int_{\theta_0}^{\theta_0+2\pi} (\eta_\tau - \eta_\theta)(\tau, \theta') \, d\theta'}
\\
& =: \Rp^\pm(\tau, \theta_0) \, H(\tau), 
\endaligned
$$ 
Recall that the functions have just been proven to be {\sl continuous} and observe
that $H$ is a positive quantity. Therefore, from the two inequalities 
$$
\Rp^-(\tau, \theta_0)  \leq \Rp^-(\tau, \theta_0) \, H(\tau), 
\qquad
\Rp^+(\tau, \theta_0)  \leq \Rp^+(\tau, \theta_0) \, H(\tau), 
$$
valid for all $\theta_0$, we deduce 
$$
\Rp^-(\tau, \theta_0) \, \Rp^+(\tau, \theta_0) (1- H(\tau))^2 \geq 0. 
$$
We recover the fact that the product $(\Rp^- \Rp^+)(\tau, \cdot)$ remains non-negative in the interval $[0,2\pi]$. 

However, if now $\Rp^+(\tau, \theta_0) = 0$ (for instance) for some $\theta_0$, then from 
$$ 
\Rp^+(\tau, \theta) \, e^{\int_{\theta_0}^{\theta} (\eta_\tau - \eta_\theta)(\tau, \theta') \, d\theta'}\Big)
= 
\int_{\theta_0}^\theta F^+(\tau, \theta') \, e^{\int_{\theta_0}^{\theta'} (\eta_\tau - \eta_\theta)(\tau, \theta'') \, d\theta''} \, d\theta'  
$$
and the fact that $F^+(\tau, \cdot)$ is non-negative, we conclude that $\Rp^+(\tau, \cdot)$ changes sign only once, 
going from non-positive to non-negative values across $\theta_0$, which is impossible for a periodic and continuous function, 
except in the special case that 
$\Rp^+(\tau, \cdot)$ is constant in $\theta$ and, in addition, $F^\pm(\tau, \cdot)$ vanishes identically. 
 
The above arguments show that the function $(\Rp^- \Rp^+)(\tau, \cdot)$ is either positive on $[0,2\pi]$ 
or else vanishes identically.  

\

\noindent{\bf 2.}  The functions $\Rp^\pm$ (and therefore $R_\tau, R_\theta$)
are continuous in both variables. Indeed, we already know that $R_{\theta\theta} \in L^\infty_\tau(L^1_\theta)$
and, in view of the equation satisfied by the function $R$,
$$
R_{\tau\tau} = R_{\theta\theta} + M^R(R) \in L^\infty_\tau(L^1_\theta).  
$$
This leads us to distinguish between three different cases for all times: 
$$
\aligned 
& \text{Case (a) : } \quad \Rp^- < 0, \quad \Rp^+ < 0, 
\\
& \text{Case (b) : } \quad  \Rp^- > 0, \quad \Rp^+ > 0, 
\\
& \text{Case (c) : } \quad  \Rp^- \equiv 0 \, \text{ or }  \, \Rp^+ \equiv 0.    
\endaligned 
$$

Suppose that, for instance, $\Rp^-(\tau, \cdot) \equiv 0$ and thus $\Rp^+(\tau, \cdot) = 2 R_\theta(\tau, \cdot)$. 
From the first part of this proof, we know that the function $\Rp^+(\tau, \cdot)$ is either positive, or negative, or identically zero. 
But, since $R_\theta(\tau, \cdot)$ is $2\pi$-periodic and {\sl continuous,} it must vanish identically. 
Consequently, both $\Rp^\pm(\tau, \cdot) \equiv 0$ and the function $R(\tau, \cdot)$ is constant in $\theta$. 
We also deduce from the expression of $F^\pm$ that 
this is the case if and only if $U_\tau, U_\theta, A_\tau, A_\theta, \rho$ vanish at the time $\tau_0$. 
Moreover, 
it is not difficult to check that the functions $U, A$ must be constant and $\rho$ be identically zero everywhere 
in their domain of definition in the future of the hypersurface $\tau = \tau_0$. The only non-trivial component the is the 
function $\eta$ which then satisfies a linear wave equation
$$
\eta_{\tau\tau} - \eta_{\theta\theta} = 0,
$$
whose solution takes the form $\eta(\tau, \theta) = f_-(\tau + \theta) + f_+(\tau - \theta)$
with $f_\pm \in W^{1,1}$.  
By setting $u_\pm := \tau \pm \theta$, the metric takes the form 
$$
g = e^{-2U} \, f_-(u_-) f_+(u_+) du_- du_+ + e^{2U} (dx + A \, dy)^2 + e^{-2U} \, R^2 \, dy^2,
$$
where the coefficients $A,U,R$ are constants. Hence, the metric is flat. 

\

\noindent{\bf 3.} 
Finally, excluding Case (c) of the proposition, 
we can compute the norm of the gradient of the function $R$ and we find 
$$
|\nabla R|^2 = e^{2(\eta- U)} \big( R_\tau^2 - R_\theta^2 \big) 
             = - e^{2(\eta- U)} \, \Rp^- \Rp^+ < 0. 
$$
From the conditions in Cases (a) and (b) we deduce that the component $R_\tau$ can not vanish
and therefore by continuity must keep a constant sign through the spacetime. 

Moreover, in Case (a) we can integrate the equation in $R$ 
$$
(\del_\tau \mp \del_\theta) \Rp^\pm = M^R \geq 0.  
$$
This yields the sup norm control of both $\Rp^\pm$, and in turn a control of both $R_\tau$ and $R_\theta$. 

\

\noindent{\bf 4.} The estimate \eqref{301} follows directly from \eqref{4-conf}, by noticing that 
the right-hand side of \eqref{4-conf} is uniformly bounded in $L^1(S^1)$. The latter indeed is easily checked
 from the fact that 
$U_\tau, U_\theta, A_\tau, A_\theta$ are uniformly bounded in $L^2(S^1)$ while $\rho$ is uniformly bounded 
in $L^1(S^1)$. 

\end{proof}

In Case (a), that is, when the spacetime is
contracting in the future direction, it will follow from our compactness framework in the next section 
that solutions in conformal coordinates do 
exist up to $\tau \to \infty$, as long as the function $R$ does not vanish. 
As in the case of smooth spacetimes, one can distinguish between two situations: 
\begin{enumerate}
\item[(a1):]  $(U,A,\eta,R,M,V)$ is a finite energy solution to the Einstein--Euler equations in 
conformal coordinates whose domain of definition $D^+_\conf$ coincides with 
$[\tau_0, +\infty) \times S^1$. 
In this case, the conformal coordinates do cover the {\sl whole} of the Cauchy development. 
\item[(a2):] $(U,A,\eta,R,M,V)$ is a finite energy solution to the Einstein--Euler equations defined 
on a region strictly contained in $[\tau_0, +\infty) \times S^1$. Here,
 $R$ 
admits a constant limit value $\Rmax$ which determines the boundary of the spacetime.
 
\end{enumerate}
%-------------------------------------------------------------------------------------------------------------------
\section{Compactness properties} 
\label{Compac}

In the previous sections, we have successfully determined the natural weak regularity assumptions 
required on the initial data and enjoyed by the solutions to the Einstein--Euler system. 
Theorem~\ref{maintheo} now follows from the compactness framework developed in~\cite{LeFloch}. 
The main property, stated below, is that 
any sequence of spacetimes satisfying the natural bounds on the geometry and fluid variables must subconverge 
in a suitably strong sense so that the limit is a spacetime with the expected regularity. 
The compactness property holds for exact solutions as well as 
approximate solutions (which should be dealt with by allowing for suitably controlled error terms). 

\begin{proposition}[Pre-compactness property for the Einstein equations under Gowdy symmetry]  
\label{compac}
Consider a sequence of spacetimes with finite energy determined by the geometric coefficients 
$U^{(a)}$, $A^{(a)}$, $\eta^{(a)}$, $R^{(a)}$ in conformal coordinates
(or $U^{(a)}$, $A^{(a)}$, $\nu^{(a)}$, $\alpha^{(a)}$ in areal coordinates) 
together with
 the (rescaled) fluid components $M^{(a)}$, $V^{(a)}$.   
Suppose that the regularity conditions stated in Definition~\ref{finit} (or Definition~\ref{finit-areal})
hold, with the corresponding 
functional norms of the initial data being uniformly bounded in the parameter $a=1,2,\ldots$. 
Then, the following pre-compactness property holds: there exists 
a subsequence (still labelled with the upper script $a$) converging in the natural functional spaces
and the limit
is a solution with finite energy to the Einstein equations. 
\end{proposition}

To conclude, we consider briefly the 
case of the Euler equations in the flat geometry, that is, in the simplified situation that 
the coupling between the matter and the geometry is neglected.  
Hence, we focus first on the fluid evolution governed by the Euler equations:  
$$ 
\aligned 
\Bigg( \mu \, {1 + k^2 \, \vv^2 \over 1 - \vv^2} \Bigg)_t 
  + \Bigg( \mu \, { (1+k^2) \, \vv \over 1 - \vv^2} \Bigg)_\theta = 0, 
\\
\Bigg( \mu \, {(1+k^2) \, \vv \over 1 - \vv^2} \Bigg)_t 
+ \Bigg( \mu \, {k^2 + \vv^2 \over 1 - \vv^2}  \Bigg)_\theta 
 = 0, 
\endaligned 
$$
where $\mu \geq 0$ and $V \in (-1,1)$ are the unknown fields, and $k \in (0,1)$ is a constant.  
Equivalently, in terms of the {\sl rescaled density} $\muh := \mu/(1-V^2)$ we have 
\be
\label{Euler-arealZ1} 
\aligned 
\Big( \muh \, (1 + k^2 \, \vv^2) \Big)_t 
  + \Big( \muh \,  (1+k^2) \, \vv \Big)_\theta = 0, 
\\
\Big( \muh \,  (1+k^2) \, \vv \Big)_t 
+ \Big( \muh \, (k^2 + \vv^2)  \Big)_\theta = 0.
\endaligned 
\ee 
According to our definition, a pair of functions $(\muh,V)$ defined on some set $\Omega \subset (t_1,t_2) \times S^1$ 
and satisfying $\muh \in L_t^\infty(L^1_\theta)$ and $V \in L^\infty$ with $V \in [-1,1]$, 
is called an entropy solution to the Euler equations \eqref{Euler-arealZ1} 
if the entropy inequalities 
$$
(\Fcal^0)_t + (\Fcal^1)_\theta \leq 0
$$
hold in the distributional sense for convex weak entropy pairs $(\Fcal^0, \Fcal^1)$. 
Note that, in the proposed formulation, the velocity may reach the limiting values $\pm 1$.

%===========================================================================================================

\section*{Acknowledgments}  

The authors are grateful to the Erwin Schr\"odinger Institute,
Vienna, where this work was completed in February 2010 during the program 
{\sl ``Quantitative Studies of Nonlinear Wave Phenomena'',}  
organized by P.C. Aichelburg, P. Bizo\'n, and W. Schlag.
This research was also supported by the A.N.R. (Agence Nationale de la Recherche)
through the grant 06-2-134423 entitled {\sl ``Mathematical Methods in General Relativity''} (Math-GR), 
and by the Centre National de la Recherche Scientifique (CNRS).

%===========================================================================================================


\begin{thebibliography}{10}
  
\bibitem{Andreasson} \auth{Andr\'easson H.,}
Global foliations of matter spacetimes with Gowdy symmetry, 
\jou{Commun. Math. Phys.} 206 (1999), 337--366.

\bibitem{AndreassonRendallWeaver} \auth{Andr\'easson H., Rendall A.D., and Weaver M.,}
Existence of CMC and constant areal time foliations in $T\sp 2$ symmetric spacetimes with Vlasov matter,
\jou{Comm. Partial Differential Equations} 29 (2004), 237--262.

\bibitem{BLSS} \auth{Barnes A.P., LeFloch P.G., Schmidt B.G., and Stewart J.M.,}
The Glimm scheme for perfect fluids on plane-symmetric Gowdy spacetimes, 
\jou{Class. Quantum Grav.} 21 (2004), 5043--5074.

\bibitem{BergerChruscielIsenbergMoncrief} \auth{Berger B.K., Chru\'sciel P.T., Isenberg J., and Moncrief V.,}
Global foliations of vacuum spacetimes with $T^2$ isometry, 
\jou{Ann. Phys.} 260 (1997), 117--148.  

\bibitem{BergerChruscielMoncrief} \auth{Berger B.K., Chru\'sciel P.T., and Moncrief V.,}
On asymptotically flat spacetimes with $G_2$-invariant Cauchy surfaces, 
\jou{Ann. Phys.} 237 (1995), 322--354. 

\bibitem{Christodoulou1} \auth{Christodoulou D.,} 
Bounded variation solutions of the spherically symmetric Einstein-scalar field equations, 
\jou{Comm. Pure Appl. Math.} 46 (1992), 1131--1220.  

\bibitem{Chrusciel} \auth{Chru\'sciel P.T.,} 
On spacetimes with $U(1) \times U(1)$ symmetric compact Cauchy surfaces,
\jou{Ann. Phys.} 202 (1990), 100--150.
  
\bibitem{ChruscielIsenbergMoncrief} \auth{Chru\'sciel P.T., Isenberg J., and Moncrief V.,} 
Strong cosmic censorship in polarized Gowdy spacetimes, 
\jou{Class. Quantum Grav.} 7 (1990), 1671--1680. 
 
\bibitem{Dafermos-book} \auth{Dafermos C.M.,}  
{\sl Hyperbolic conservation laws in continuum physics,}  
Grundlehren Math. Wissenschaften Series 325, 3rd edn., Springer, New York, 2010. 

\bibitem{DafermosRendall} \auth{Dafermos M. and Rendall A.D.,}
Strong cosmic censorship for $T^2$-symmetric cosmological spacetimes with collisionless matter,
Preprint gr-qc/0610075. 

\bibitem{EardleyMoncrief} \auth{Eardley D. and Moncrief V.,}
The global existence problem and cosmic censorship in general relativity, 
\jou{Gen. Relat. Grav.} 13 (1981), 887--892.  

\bibitem{Geroch1} \auth{Geroch R.,}
A method for generating solutions of Einstein's equations, 
\jou{J. Math. Phys.} 12 (1971), 918--924.

\bibitem{Geroch2} \auth{Geroch R.,}
A method for generating new solutions of Einstein's equations. II, 
\jou{J. Math. Phys.} 13 (1972), 394--404. 

\bibitem{Gowdy} \auth{Gowdy R.,} 
Vacuum spacetimes with two-parameter spacelike isometry groups and compact invariant hypersurfaces: 
topologies and boundary conditions, 
\jou{Ann. Phys.} 83 (1974), 203--241. 

\bibitem{GroahTemple} \auth{Groah J. and Temple B.,} 
{\sl Shock-wave solutions of the Einstein equations: existence and consistency by a locally inertial Glimm scheme}, Mem. Am. Math. Soc.~172, No. 813 (2004).
 
\bibitem{HawkingPenrose} \auth{Hawking S.W. and Penrose R.,}  
The singularities of gravitational collapse and cosmology, 
\jou{Proc. Roy. Soc.} A314 (1970), 529--548.   

\bibitem{IsenbergMoncrief} \auth{Isenberg J. and Moncrief V.,}
Asymptotic behavior of the gravitational field and the nature of singularities in Gowdy spacetimes, 
\jou{Ann. Phys.} 99 (1990), 84--122. 

\bibitem{IsenbergWeaver} \auth{Isenberg J. and Weaver M.,}
On the area of the symmetry orbits in $T^2$ symmetric spacetimes, 
\jou{Class. Quantum Grav.} 20 (2003), 3783--3796.  

\bibitem{LeFloch-book} \auth{LeFloch P.G.,} 
{\sl Hyperbolic systems of conservation laws: The theory of classical and nonclassical shock waves}, 
Lectures in Mathematics, ETH Z\"urich, Birkh\"auser, 2002.

\bibitem{LeFloch} \auth{LeFloch P.G.,} 
Compressible fluids in curved spacetimes. In preparation. 

\bibitem{LeFlochMardare} \auth{LeFloch P.G. and Mardare C.,}
Definition and weak stability of spacetimes with distributional curvature, 
\jou{Port. Math.} 64 (2007), 535--573.

\bibitem{LeFlochStewart1} \auth{LeFloch P.G. and Stewart J.M.,}
Shock waves and gravitational waves in matter spacetimes with Gowdy symmetry, 
\jou{Port. Math.} 62 (2005), 349--370. 

\bibitem{LeFlochStewart2} \auth{LeFloch P.G. and Stewart J.M.,}
The characteristic initial value problem for plane symmetric spacetimes with weak regularity, 
Class. Quantum Grav. (2011).
% Preprint ArXiv:1004.2343 (April 2010) . 

\bibitem{Moncrief} \auth{Moncrief V.,}
Global properties of Gowdy spacetimes with $T^3 \times \RR$ topology, 
\jou{Ann. Phys.} 132 (1981), 87--107. 

\bibitem{Penrose} \auth{Penrose R.,}
Gravitational collapse and spacetime singularities, 
\jou{Phys. Rev. Lett.} 14 (1965), 57--59. 

\bibitem{Rendall1} \auth{Rendall A.D.,} 
Cosmic censorship and the Vlasov equation, 
\jou{Class. Quantum Grav.} 9 (1992), 99--104. 

\bibitem{Rendall-crush} \auth{Rendall A.D.,} 
Crushing singularities in spacetimes with spherical, plane, and hyperbolic symmetry, 
\jou{Class. Quantum Grav.} 12 (1995), 1517--1533. 

\bibitem{Rendall2} \auth{Rendall A.D.,} 
Existence of constant mean curvature foliations in spacetimes with two-dimensional local symmetry,
\jou{Commun. Math. Phys.} 189 (1997), 145--164.

\bibitem{rendall08} \auth{Rendall A.D.,} 
{\em Partial differential equations in general relativity,}
Oxford University Press, Oxford, 2008.

\bibitem{rendallstahl} \auth{Rendall A. D. and St{\aa hl} F.,} 
Shock waves in plane symmetric spacetimes,
\jou{Commun. Partial Differential Equations} 33 (2008), 2020--2039.

\bibitem{Ringstrom1}\auth{Ringstr\"om H.,} 
Curvature blow-up on a dense subset of the singularity in $T^3$-Gowdy, 
\jou{J. Hyper. Diff. Equa.} 2 (2005), 547--564. 

\bibitem{Ringstrom2}\auth{Ringstr\"om H.,} 
Strong cosmic censorship in $T^3$-Gowdy spacetimes, 
\jou{Ann. Math.} 170 (2009), 1181--1240. 

\bibitem{Smulevici} \auth{Smulevici J.,}
Strong cosmic censorship for $T^2$-symmetric spacetimes with positive cosmological constant and matter, 
\jou{Ann. Henri Poincar\'e} 9 (2009), 1425--1453.

\bibitem{wainwright} \auth{Wainwright J. and Ellis G.F.R.,}
\jou{Dynamical systems in cosmology,} Cambridge University Press, Cambridge, 1997.

\bibitem{weaver} \auth{Weaver M.,} 
On the area of the symmetry orbits in $T^2$-symmetric spacetimes with Vlasov matter,
\jou{Class. Quantum Grav.} 21 (2004), 1079--1098.

\end{thebibliography}
\end{document}